\PassOptionsToPackage{prologue,svgnames,table}{xcolor}
\PassOptionsToPackage{numbers, sort&compress}{natbib}
\documentclass[sigconf,nonacm]{acmart}

\usepackage[T1]{fontenc}
\usepackage[utf8]{inputenc}
\usepackage{textgreek}

\makeatletter
\begingroup\endlinechar=-1\relax
       \everyeof{\noexpand}%
       \edef\x{\endgroup\def\noexpand\homepath{%
         \@@input|"kpsewhich --var-value=HOME" }}\x
\makeatother
\def\overleafhome{/tmp}

\usepackage{import}

\ifx\homepath\overleafhome
  \newrobustcmd{\toolspath}[0]{latex-tools-overleaf/}
\else
  \newrobustcmd{\toolspath}[0]{latex-tools-overleaf/}
\fi

\usepackage{etoolbox}

\newtoggle{AlgsNoEnd}

\toggletrue{AlgsNoEnd}

\newif\iftr     % Full arXiv technical report
\newif\ifconf   % Size-constrainted Submission to a conf or journal
\newif\ifnonb   % Non blind submission

\usepackage[utf8]{inputenc}
\newcommand{\code}[1]{\texttt{#1}}
\usepackage[binary-units=true]{siunitx}
\usepackage{xspace}
\usepackage[svgnames]{xcolor}

\usepackage{collab}
\usepackage[numbers,sort&compress]{natbib}

\usepackage{soul}
\soulregister\cite7
\soulregister\autoref7
\soulregister\code7
\soulregister\toolname7
\soulregister\ref7
\soulregister\pageref7
\usepackage{hhline}
\definecolor{lightyellow}{RGB}{250, 250, 180}
\definecolor{HLYELLOW}{RGB}{240, 127, 0}
\definecolor{hlyellow}{RGB}{240, 127, 0}
\sethlcolor{lightyellow}

\usepackage{algorithm}
\usepackage{algpseudocode}
\usepackage{pifont}
\usepackage{amsmath}
\algdef{SE}[FOREACHIN]{ForEachIn}{EndForEachIn}[2]{\algorithmicfor\ \textbf{each}\ #1\ \textbf{in}\ #2\ \textbf{do}}{\algorithmicend\ \algorithmicfor}%
\algdef{SE}[FOREACHP]{ForEachP}{EndForEachP}[1]{\algorithmicfor\ \textbf{each}\ #1\ \textbf{in parallel do}}{\algorithmicend\ \algorithmicfor}%
\algdef{SE}[FOREACHINP]{ForEachInP}{EndForEachInP}[2]{\algorithmicfor\ \textbf{each}\ #1\ \textbf{in}\ #2\ \textbf{in parallel do}}{\algorithmicend\ \algorithmicfor}%
\algdef{SE}[FORMOD]{ForMod}{EndForMod}[3]{\algorithmicfor\ #1\ \textbf{from}\ #2\ \textbf{to}\ #3\ \textbf{do}}{\algorithmicend\ \algorithmicfor}%
\algdef{SE}[FORMOD]{ForModS}{EndForModS}[4]{\algorithmicfor\ #1\ \textbf{from}\ #2\ \textbf{to}\ #3\ \textbf{step}\ #4\ \textbf{do}}{\algorithmicend\ \algorithmicfor}%
\algdef{SE}[FORMODP]{ForModPS}{EndForModPS}[3]{\algorithmicfor\ #1\ \textbf{from}\ #2\ \textbf{to}\ #3\ \textbf{in parallel do}}{\algorithmicend\ \algorithmicfor}%
\algdef{SE}[FORMODP]{ForModP}{EndForModP}[4]{\algorithmicfor\ #1\ \textbf{from}\ #2\ \textbf{to}\ #3\ \textbf{step} #4\ \textbf{in parallel do}}{\algorithmicend\ \algorithmicfor}%

\iftoggle{AlgsNoEnd} {
  \algtext*{EndFunction}% Remove "end while" text
  \algtext*{EndIf}% Remove "end if" text
  \algtext*{EndFor}% Remove "end if" text
  \algtext*{EndForEachIn}% Remove "end if" text
  \algtext*{EndForEachP}% Remove "end if" text
  \algtext*{EndForEachInP}% Remove "end if" text
  \algtext*{EndForModP}% Remove "end if" text
  \algtext*{EndForModPS}% Remove "end if" text
  \algtext*{EndForModS}% Remove "end if" text
}{}

\algdef{SE}[VARIABLES]{Variables}{EndVariables}{\algorithmicvariables}
  {\algorithmicend\ \algorithmicvariables}
  \algnewcommand{\algorithmicvariables}{\textbf{global}}
\algnewcommand{\LineComment}[1]{\State \(\triangleright\) #1}
\algnewcommand{\And}{\textbf{and}\xspace}

\usepackage{tikz}
\usepackage{pifont}
\usetikzlibrary{shapes}

\DeclareRobustCommand*\circledColor[2]{\tikz[baseline=(char.base)]{
    \node[shape=circle,fill=#2,draw=#2,inner sep=1pt] (char) {\textcolor{white}{\small\textbf{#1}}};}}

\DeclareRobustCommand*\nodeColor[3]{%
  \tikz[baseline=(char.base)]{%
    \node[shape=#3,fill=#2,draw=#2,inner sep=0pt] (char) {\textcolor{white}{\small\textbf{#1}}};%
  }%
}

\usepackage{listings}
\newfloat{codeblock}{H}{myc}
\definecolor{darkblue}{rgb}{0,0,.6}
\definecolor{darkred}{rgb}{.6,0,0}
\definecolor{darkgreen}{rgb}{0,.5,0}
\definecolor{red}{rgb}{.98,0,0}
\definecolor{gray}{rgb}{.6,.6,.6}
\definecolor{newgreen}{RGB}{169,209,142}
\definecolor{newpurple}{RGB}{237,134,254}
\definecolor{neworange}{RGB}{244,177,131}
\definecolor{newyellow}{RGB}{255,217,102}
\collabAuthor{copik}{blue}{Marcin}
\collabAuthor{lex}{violet}{lex}
\collabAuthor{htor}{magenta}{Torsten}

\usepackage{subcaption}
\usepackage{listings}
\usepackage{tabularx}
\usepackage{adjustbox}
\usepackage{makecell}
\usepackage{nccmath}

\usepackage{pifont}% http://ctan.org/pkg/pifont
\newcommand{\cmark}{\ding{51}}%
\newcommand{\mmark}{\ding{83}}%
\newcommand{\xmark}{\ding{55}}%

\newcommand{\minitab}[2][l]{\begin{tabular}{#1}#2\end{tabular}}
\usepackage{array}
\newcolumntype{C}[1]{>{\centering\arraybackslash}p{#1}}

\usepackage{enumitem}

\usepackage{url}
\usepackage{booktabs,tabularx,multirow}

\usepackage{textcomp}
\definecolor{darkgreen}{rgb}{0.0, 0.5, 0.0}
\definecolor{bananayellow}{rgb}{1.0, 0.88, 0.21}
\definecolor{codegreen}{rgb}{0,0.6,0}
\definecolor{codegray}{rgb}{0.5,0.5,0.5}
\definecolor{codepurple}{rgb}{0.58,0,0.82}
\definecolor{backcolour}{rgb}{0.95,0.95,0.92}
\definecolor{PigRed}{RGB}{255, 0, 0}
\definecolor{gold}{RGB}{191,144,0}
\lstdefinestyle{zookeeper}{
	backgroundcolor=\color{backcolour},
	commentstyle=\color{codegreen},
	keywordstyle=\color{magenta},
	numberstyle=\tiny\color{codegray},
	stringstyle=\color{codepurple},
  morecomment=[l]{//},
  morestring=[d]{"},
	basicstyle=\ttfamily\footnotesize,
	breakatwhitespace=false,
	breaklines=true,
	captionpos=b,
	keepspaces=true,
	numbers=left,
	numbersep=5pt,
	showspaces=false,
	showstringspaces=false,
	showtabs=false,
	tabsize=2,
  numbers=left,
  keywords=[1]{
    create,setData,getData,exists
  },
  keywordstyle=[1]\color{magenta},
  keywords=[2]{
    NULL, EPHEMERAL, SEQUENTIAL, NONE
  },
  keywordstyle=[2]\color{PigRed}
}

\usepackage{array}
\newcolumntype{L}[1]{>{\raggedright}m{#1}}

\usepackage[fixed]{fontawesome5}
\DeclareSIUnit{\microsecond}{\SIUnitSymbolMicro s}

\newcommand{\toolname}{FaaSKeeper\xspace{}}
\newcommand{\reqCount}{nine\xspace{}}

\usepackage[many]{tcolorbox}
\tcbuselibrary{skins}
\usetikzlibrary{calc}

\definecolor{textnewBeige}{HTML}{d5bdaf} %{e3d5ca}
\definecolor{boxOliveGreen}{HTML}{a5a58d} %{e3d5ca}
\definecolor{myblue}{RGB}{0,163,243}
\definecolor{textnewGreen}{HTML}{66AE3E}
\definecolor{bgnewGreen}{HTML}{EBF4DE}
\newtcolorbox{summaryBox}[1]{
  enhanced,
  skin=bicolor,
  arc=0pt,
	left=0pt,
	right=0pt,
	top=0pt,
	bottom=0pt,
  coltitle=white,
  colframe=#1,
  colback=#1!20,
  colbacklower=white
}

\begin{document}
%-------------------------------------------------------------------------------

\title{FaaSKeeper: Learning from Building Serverless Services with ZooKeeper as an Example}

\author{Marcin Copik}
\affiliation{%
  \institution{ETH Zurich}
  \country{Switzerland}
}
\email{marcin.copik@inf.ethz.ch}

\author{Alexandru Calotoiu}
\affiliation{%
  \institution{ETH Zurich}
  \country{Switzerland}
}
\email{alexandru.calotoiu@.inf.ethz.ch}

\author{Pengyu Zhou}
\affiliation{%
  \institution{University of Toronto}
  \country{Canada}
}
\email{ericpengyu.zhou@mail.utoronto.ca}

\author{Konstantin Taranov}
\affiliation{%
  \institution{Microsoft}
  \country{Switzerland}
}
\email{kotaranov@microsoft.com}

\author{Torsten Hoefler}
\affiliation{%
  \institution{ETH Zurich}
  \country{Switzerland}
}
\email{htor@inf.ethz.ch}




%-------------------------------------------------------------------------------
\begin{abstract}
  FaaS (Function-as-a-Service) revolutionized cloud computing by replacing persistent virtual machines with dynamically allocated resources.
  This shift trades locality and statefulness for a pay-as-you-go model more suited to variable and infrequent workloads.
  However, the main challenge is to adapt services to the serverless paradigm while meeting functional, performance, and consistency requirements.
  %
  %In this work, we redesign ZooKeeper, a centralized coordination service with a safe and wait-free consensus mechanism, to harness the benefits of serverless computing.
  In this work, we push the boundaries of FaaS computing by designing a serverless variant of ZooKeeper, a centralized coordination service with a safe and wait-free consensus mechanism.
  We define synchronization primitives to extend the capabilities of scalable cloud storage
  and outline a set of requirements for efficient computing with serverless.
  In FaaSKeeper, the first coordination service built
  on serverless functions and cloud-native services,
  we explore the limitations of serverless offerings and propose improvements essential for complex and latency-sensitive applications.
  We share serverless design lessons based on our experiences of implementing a ZooKeeper model deployable to clouds today.
  FaaSKeeper maintains the same consistency guarantees and interface as ZooKeeper,
  with a serverless price model that lowers costs up to 110-719x on infrequent workloads.
  %
%-------------------------------------------------------------------------------
\end{abstract}

\maketitle

\vspace{-1em}
{\small\noindent\textbf{FaaSKeeper implementation:} \url{https://github.com/spcl/faaskeeper}}

{\small\noindent\textbf{FaaSKeeper Artifact:} \url{https://github.com/spcl/faaskeeper-paper-artifact}}

{\small\noindent\textbf{Paper version published at ACM HPDC 2024:} \url{https://doi.org/10.1145/3625549.3658661}}

%-------------------------------------------------------------------------------
\section{Introduction}

\setlength{\tabcolsep}{3pt}
\begin{table}[t!]\centering
\footnotesize
%\begin{tabularx}{\linewidth}{L{0.05cm}lll@{}}
% @{} - suppress the column separation space
\begin{tabularx}{\linewidth}{@{}L{1em}lll@{}}
%\toprule
& \textbf{ZooKeeper} & \textbf{Cloud Storage} & \textbf{FaaSKeeper}\\
%\midrule
\hline
  \faIcon{angle-double-up} & Semi-automatic, $\ge 3$ VMs & \textbf{Automatic} & \textbf{Automatic} \\
  % Scaling Down
  \faIcon{angle-double-down} & Not possible. & \textbf{Only storage fees}& \textbf{Only storage fees}\\
  \faIcon{dollar-sign} & Pay upfront & \textbf{Pay-as-you-go} & \textbf{Pay-as-you-go} \\
  % High Availability and reliability
  \faIcon{shield-alt} & Depends on cluster size & \textbf{Cloud SLA} & \textbf{Cloud SLA}\\
  % Consistency
  \faIcon{sync-alt} & \textbf{Linearized writes} & Strong consistency & \textbf{Linearized writes}\\
  % Push Notifications
  \faIcon{envelope} & \textbf{Watch events} & None & \textbf{Watch events}\\
  % Concurrency
  \multirow{2}{*}{\faIcon{compress-arrows-alt}} & \textbf{Sequential nodes}, & \multirow{2}{*}{Conditional updates.} & \textbf{Sequential nodes},\\
                               & conditional updates &  & conditional updates\\
  % Failure
  \faIcon{bug} & \textbf{Ephemeral nodes} & None & \textbf{Ephemeral nodes}\\
  \bottomrule
\end{tabularx}
%\end{tabular}
\caption{\toolname{} combines the best features of cloud storage: scale--to--zero ({\hspace{-0.25em}\small\faIcon{angle-double-up}\hspace{-0.75em}\faIcon{angle-double-down}\hspace{-0.25em}}) and reliability ({\small\hspace{-0.25em}\faIcon{shield-alt}\hspace{-0.25em}}), with ZooKeeper's consistency ({\small\hspace{-0.25em}\faIcon{sync-alt}\hspace{-0.25em}}), push notifications ({\small\hspace{-0.25em}\faIcon{envelope}\hspace{-0.25em}}), and support for concurrency and fault tolerance ({\hspace{-0.25em}\small\faIcon{compress-arrows-alt}\hspace{-0.25em}\faIcon{bug}\hspace{-0.25em}}).}
\label{tab:fk_zk_comparison}
\vspace{-3em}
\end{table}

% what changed in the field - serverless
FaaS is a new paradigm that combines elastic and on-demand resource allocation
with an abstract programming model.
In FaaS, the cloud provider invokes stateless functions,
freeing the user from managing the software and hardware resources.
Flexible resource management and a pay-as-you-go billing %system
help with the problem of low server utilization
caused by resource overprovisioning for the peak workload
~\cite{10.1145/1721654.1721672,kaplan2008revolutionizing,6253523}.
These improvements come at the cost of performance and reliability:
functions are not designed for high-performance applications and
require storage to support state and communication.
However, stateful applications can benefit from
serverless services~\cite{10.1145/3476886.3477510},
and even databases adapt on-demand offerings
to handle
infrequent workloads more
efficiently~\cite{awsDynamoDBServerless,azureCosmosDBServerless,kassandraServerless}.

Apache ZooKeeper~\cite{10.5555/1855840.1855851} is a prime example of a system
that has been widely adopted by many applications but is not available
as a serverless service.
ZooKeeper provides a coordination service for distributed applications to
control the shared state and guarantee data consistency and availability.
Compared to key-value storage, ZooKeeper adds semantics of total order with linearizable writes,
atomic updates, and ordered push notifications (Table~\ref{tab:fk_zk_comparison}).
%

%Modern
Cloud services are expected to match the temporal and geographical
variability of production workloads~\cite{4362193,10.1145/2987550.2987561,10.1145/3132747.3132772}.
Workloads are often bursty and experience rapid changes:
maximum system utilization can be multiple times higher than even the 99th percentile~\cite{10.5555/1855807.1855835,4362193}.
However, the static ZooKeeper architecture %and allocation of compute resources
make the readjustment to the workload difficult,
and ZooKeeper is often underutilized in practical deployments (Section~\ref{sec:zookeeper_utilization}).
Even when ZooKeeper is co-located as a part of a larger system, it still contributes
to the overprovisioning of resources for the peak workload.
Serverless applications built on cloud storage could adapt to diurnal changes in workload
and handle thousands of requests at a lower cost.
A \emph{serverless} service with the same consistency as ZooKeeper would offer the
opportunity to consolidate variable workloads, helping users and cloud operators increase efficiency.
Unfortunately, the path to serverless for such distributed applications
is unclear due to the restricted and vendor-specific nature of FaaS.

In this work, \textbf{we chart the path needed to build a complex serverless service} --- serverless ZooKeeper.
We choose ZooKeeper \textbf{because} it is a complex, \textbf{reliable} service, and therefore challenges both the capabilities and the limitations of inherently unreliable FaaS systems,
which lack fast communication channels, ordering, and statefulness.
First, we decouple the system from the application state and
compute from storage tasks~\cite{10.1145/2445583.2445588,10.1145/3318464.3386129,10.1145/3448016.3457560},
Similar to past results in building a database on cloud storage~\cite{10.1145/1376616.1376645},
we build a serverless architecture on top of auto-scalable storage,
focusing on a \emph{cloud-native} design that requires no user-managed and custom solutions,
but is instead deployable to clouds available today (Section~\ref{sec:fk_to_zk}).
We focus on the semantics of building components and abstract away differences in interfaces
and services, helping design \emph{cloud-agnostic} systems that are portable
between clouds (Section~\ref{sec:faaskeeper-optimizations}).

Finally, we introduce and evaluate \textbf{\toolname{}}, the first coordination service with a serverless scaling and billing model.
In \toolname{}, we combine the best of two worlds: the ordered transactions and active
notifications of ZooKeeper with cloud storage's elasticity
and high reliability (Table~\ref{tab:fk_zk_comparison}).
We implement ZooKeeper's model and API in FaaS, with a prototype of the provider-agnostic system on AWS and GCP (Section~\ref{sec:faaskeeper-optimizations}),
demonstrating consensus in a serverless application on top of consistent and replicated cloud storage.
\toolname{} offers a pay-as-you-go cost model while upholding consistency
and ordering properties (Section~\ref{sec:faaskeeper_properties}).
Another goal of this paper is to demonstrate the fundamental trade-offs of moving data-intensive services to new cloud paradigms such as serverless.
We explore design choices in storing and updating cloud data, proving the cost and resource efficiency of serverless while enumerating limitations of current serverless offerings 
(Section~\ref{sec:discussion}).

In summary, we make the following contributions:
\begin{itemize}[noitemsep,topsep=0pt,leftmargin=2.5em]

  \item Exploration of challenges and limitations in serverless and lessons
  for designing cloud-native services with synchronization, message ordering, and event-based communication.
  \item The first complex serverless solution
  that offers the same level of service as its IaaS counterpart without provisioning.

  \item An API-compatible implementation of the ZooKeeper consistency model
    that achieves up to 60 times lower costs
    against the smallest ZooKeeper deployment.

\end{itemize}

%-------------------------------------------------------------------------------

%-------------------------------------------------------------------------------

\vspace{-1em}
\section{Background and Motivation}
\label{sec:background}
While serverless systems differ between cloud providers, they can represented as 
fundamental building blocks needed to design serverless services (Section~\ref{sec:faaskeeper_serverless_components}).
These are necessary to implement in serverless the distributed coordination model 
of ZooKeeper (Section~\ref{sec:faaskeeper_zookeeper}).

\subsection{Serverless Components}
\label{sec:faaskeeper_serverless_components}
\begin{figure}[t!]
	\centering
  \includegraphics[width=\linewidth,keepaspectratio]{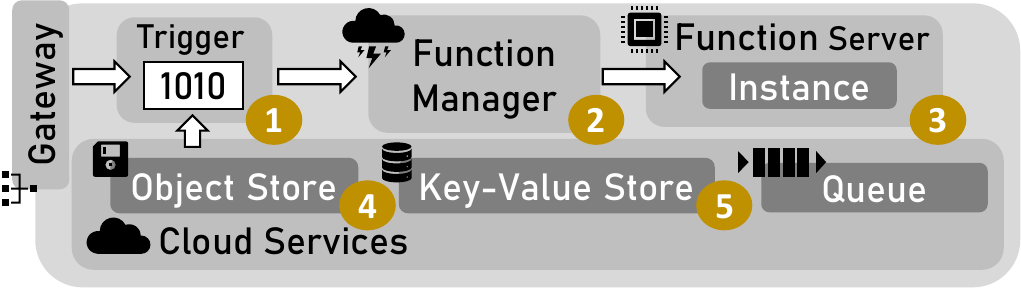}
  %\vspace{-2.0em}
  \caption{A high-level overview of a FaaS platform.}
  %\vspace{-2em}
  \label{fig:faas_model}
\end{figure}

Serverless functions replace persistent virtual machines with elastic and dynamic execution
of fine-grained tasks (Figure~\ref{fig:faas_model}).
%complementing computing and storage solutions in the cloud.
%
The management of the software and hardware stack becomes the sole responsibility
of the cloud provider, and users are charged only for the time and resources
consumed during the function execution (\emph{pay-as-you-go}).
In place of cloud resource management and orchestration systems,
functions offer various \emph{triggers} to process internal cloud events
and external REST requests (\circledColor{1}{gold}).
A function scheduler (\circledColor{2}{gold}) routes the invocation to a selected server~\cite{246288},
and the function executes within an isolated sandbox on a multi-tenant server (\circledColor{3}{gold}).
The cloud scheduler aims to increase performance by reusing sandboxes,
since \emph{warm} execution in an existing container is faster than
\emph{cold} invocations that wait for sandbox allocation.

\textbf{Cloud Storage.}
Cloud operators offer storage solutions that differ in elasticity, costs,
reliability, and performance.

\noindent\emph{Object.}
Object storage is designed to store large amounts of data for a long time
while providing high throughput and durability (\circledColor{4}{gold}).
The cloud operator manages replication across multiple instances in physically
and geographically separated data centers, providing high availability and reliability.
Modern object stores offer strong consistency on read operations~\cite{10.1145/2043556.2043571},
guaranteeing that successful writes are immediately visible to other clients.
The billing model is linear in the data amount and the number of performed operations.

\noindent\emph{Key-Value.}
Nonrelational databases are common in serverless applications (\circledColor{5}{gold}),
and they also offe serverless billing where the costs depend 
only on the stored data and operations performed.
In addition to strong consistency, read operations can be executed with
eventual consistency~\cite{vogels2009eventually},
trading consistency for lower costs, improved latency, and higher availability.
They can offer optimistic concurrency with \emph{conditional} updates
that apply atomic operations to existing attributes.

\noindent\emph{Other.}
FaaS can employ additional storage systems, but these often
introduce resource provisioning.
\emph{Ephemeral} storage~\cite{216007, 10.5555/3291168.3291200} is
designed to meet serverless requirements for scalability and flexibility.
In-memory caches bring lower latency and are being adapted to serverless scalability~\cite{246184,upstash}.

\textbf{Functions}
We specify three distinct classes of functions that are necessary to implement
a serverless application or a microservice and have divergent interfaces and fault-tolerance models
--- their semantics express different programming language constructs.
A \textbf{free function } is not bound to any cloud resource and is invoked via an API request.
It can be invoked synchronously anytime, by anyone, from any location,
as long as the authorization succeeds
and the constraints on concurrent invocations are satisfied.
Free functions express the semantics of \emph{remote procedure calls}~\cite{nelson1981remote}.
The event-driven programming paradigm is implemented by providing \textbf{event functions} to react
to specific changes in cloud storage, databases, or queues.
There, API requests are replaced by sending a message to a queue that triggers the function.
Furthermore, using such a proxy allows coalescing many invocations into a larger batch
and preserving their internal ordering.
From the client's point of view, sending a message to a queue-triggered
function replaces passing requests over a TCP connection to a
server of a non-serverless service.
We expect each trigger to have configurable batching and concurrency of invocations,
as the former improves throughput and the latter is essential to ensure FIFO order.
Semantically, we interpret these functions as remote \emph{asynchronous callbacks}
to events.

Functions can be launched to perform regular routines such as garbage collection
and detecting system faults.
\textbf{Scheduled functions} are the serverless counterpart of a \code{cron} job
in Unix-based operating systems.
In the event of an unexpected failure, the cloud should provide a retry policy
with a finite number of repetitions.
Users should be notified of repeated errors to detect system-wide
failures, even when they do not directly control such functions.

\textbf{Synchronization Primitives}
Functions operating in parallel require fundamental synchronization primitives
to safely modify a global state~\cite{10.5555/2385452}, as it is the case in \toolname{}.
In serverless, such primitives operate on storage instead of shared memory.

A \textbf{timed lock} extends a regular lock with a limited holding time, similarly to leases~\cite{gray1989leases}.
It is a necessary feature to prevent a system-wide deadlock caused by a failure of an ephemeral function.
Lock operations are submitted with a user timestamp.
The lock is \emph{acquired} if no timestamp is present or when the difference between the existing
and new timestamp is greater than a predefined maximum time.
To prevent accidental overwriting after losing the lock, each update to a locked resource
compares the stored timestamp with the user value.
The lock \emph{release} removes the timestamp.
An \textbf{atomic counter} supports single-step updates
while \textbf{atomic list} provides safe expansion and truncation.

\begin{figure}[t!]
  \centering
  \begin{subfigure}{\linewidth}
  
  \includegraphics[width=\textwidth,keepaspectratio]{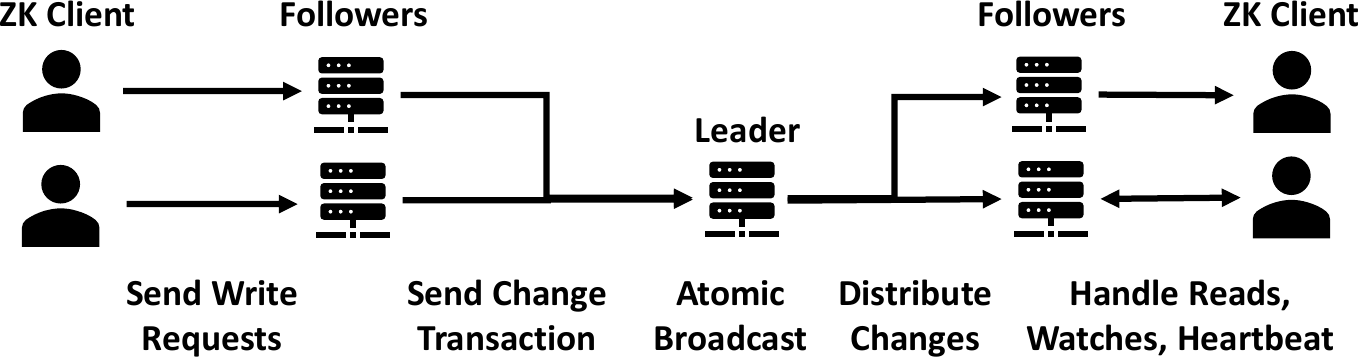}
    \vspace{-1em}
    \caption{ZooKeeper processes requests by a set of servers \emph{following} a leader.}
    \label{fig:new_faaskeeper_zookeeper}
  
  \end{subfigure}
	\\
  \begin{subfigure}{\linewidth}
  \includegraphics[width=\textwidth,keepaspectratio]{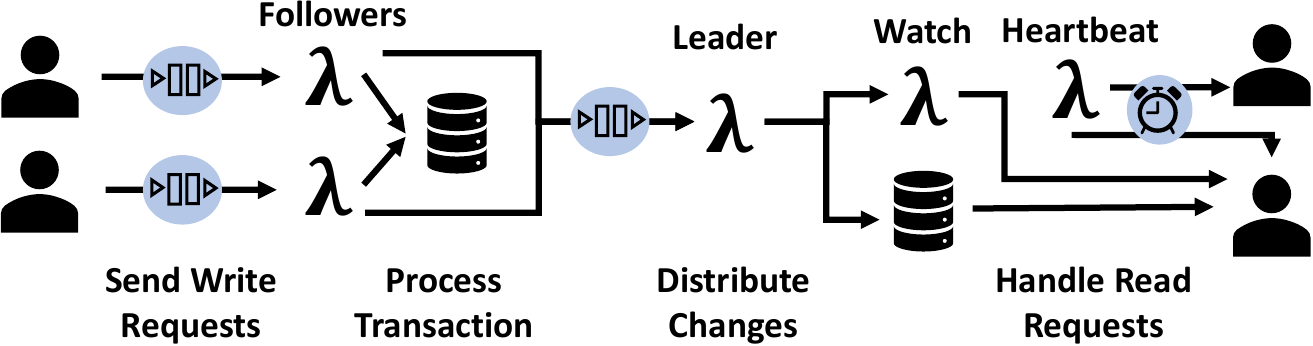}
   \caption{FaaSKeeper maps ZooKeeper to stateless functions and storage.}
    \label{fig:new_faaskeeper_zookeeper_fk}
  %\vspace{-1.0em}
  \end{subfigure}
  \vspace{-1em}
	\caption{
    \textbf{From ZooKeeper servers to functions in \toolname{}.}
	}
  %\vspace{-1.75em}
  \label{fig:new_faaskeeper}
\end{figure}

\subsection{ZooKeeper}
\label{sec:faaskeeper_zookeeper}
ZooKeeper guarantees data persistence and high read performance by allocating replicas of the entire
system on multiple servers~\cite{10.5555/1855840.1855851,10.5555/2904421,zookeeperDocs}.
ZooKeeper ensemble consists of \emph{followers} and an elected leader (Figure~\ref{fig:new_faaskeeper_zookeeper}),
whose roles are processing write requests
with the help of the ZAB atomic broadcast protocol~\cite{5958223}.
The smallest ZooKeeper deployment uses three servers, where two are required to accept
a change and failure of one can be tolerated.
While adding more servers increases reliability, it hurts write performance.

In ZooKeeper, changing the deployment size
involved \emph{rolling restarts}, a manual and error-prone process~\cite{zookeeperDynReconf}.
%
%https://tech.smartling.com/self-healing-apache-zookeeper-cluster-470b248ccb12
%https://www.credera.com/insights/how-to-automate-zookeeper-in-aws/
%https://stackoverflow.com/questions/44815521/multi-region-apache-kafka-and-zookeeper-in-autoscaling-group-in-aws
%https://zookeeper.apache.org/doc/current/zookeeperReconfig.html#sc_reconfig_modifying
%
While it has been later enhanced with dynamic reconfiguration~\cite{181005},
it still requires manual effort~\cite{zookeeperDynReconf},
and reconfiguration causes significant performance degradation when deploying
across geographical regions~\cite{10.1145/2987550.2987561}.

ZooKeeper splits the responsibilities between the client library, follower servers, and the elected leader.
User data is stored in \emph{nodes}, which create a tree structure with parents and children.
Clients send requests to a server through a session mechanism that
guarantees the
FIFO order of processing requests, achieved over reliable and fast TCP connections.
Read requests are resolved using a local data replica, while write operations are forwarded to the leader.
The leader updates nodes, manages the voting process, and propagates changes to other servers.
ZooKeeper defines the order of transactions with a monotonically increasing counter \emph{zxid}.
While requests from a single client cannot be reordered,
the order of operations between different sessions is not specified.
Clients register \emph{watches} on a node to receive a push notification when that node changes.
Finally, clients exchange heartbeat messages with a server to keep the session alive.

\textbf{Consistency.} %Model.}
ZooKeeper implements sequential consistency guarantees with four main requirements (\nodeColor{Z}{black}{ellipse}).
All operations of a single client are executed \textbf{atomically} (\nodeColor{Z1}{black}{ellipse}),
in FIFO order, and writes are \textbf{linearized} %~\cite{10.1145/78969.78972}
(\nodeColor{Z2}{black}{ellipse}).
The order of transactions is \emph{total} and \emph{global}. Thus, all clients have
a \textbf{single system image} (\nodeColor{Z3}{black}{ellipse}) and observe the same order of updates.
Watches ensure that clients know about a change before observing subsequent updates since notifications are \textbf{ordered}
with read and write operations (\nodeColor{Z4}{black}{ellipse}).
Formal definitions can be found in the Appendix~\ref{sec:appendix_zookeeper}.

%-------------------------------------------------------------------------------

%-------------------------------------------------------------------------------
\section{From ZooKeeper to FaasKeeper}

\label{sec:fk_to_zk}

\begin{figure}[t!]
	\centering
  \includegraphics[width=\linewidth,keepaspectratio]{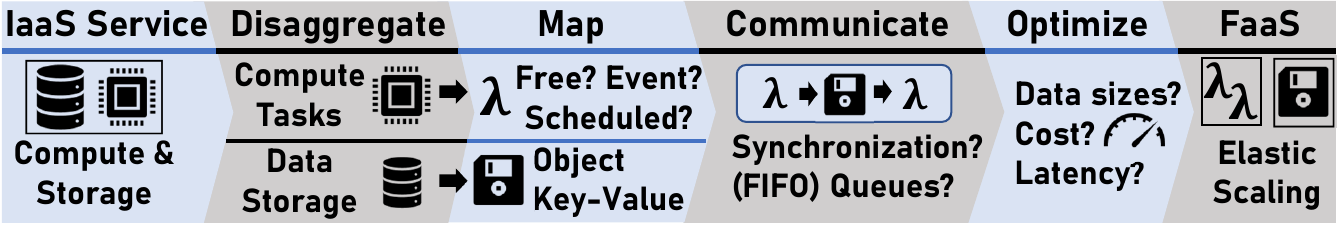}
  \vspace{-1.0em}
	\caption{
    \textbf{Workflow for designing a serverless service. The decoupled compute and storage are connected to cloud services, and later optimized for cost and performance (Section~\ref{sec:faaskeeper-optimizations}).}
	}
  \label{fig:wf}
\end{figure}

The design and implementation of ZooKeeper are incompatible with the serverless paradigm and require us to build a reliable service on top of a fundamentally unreliable FaaS foundation.
Therefore, we designed \toolname{} from the ground up to replicate the complex ZooKeeper functionality and overcome the inherent challenges of the serverless world. 
We follow a general workflow for turning an IaaS system into FaaS (Figure~\ref{fig:wf}):
disaggregate compute and data,
replace servers with cloud storage and stateless functions, 
%map them onto cloud storage and functions,
and let the new components communicate.
By decomposing the system into separate cloud services -
functions, storage, and queues (Figure~\ref{fig:new_faaskeeper_zookeeper_fk})
- we adjust the resource consumption to throughput and shut down processing
instances when they are no longer needed.
While the transition of many applications and microservices to serverless might be straightforward,
ZooKeeper has non-trivial ordering and data visibility requirements.
Thus, we design custom synchronization, queue communication, and new algorithms to implement ZooKeeper's data model in the serverless world.

We describe each component of \toolname{} by following the path of a client performing a write operation, then discuss all additional components.
We map the computational logic of \emph{follower} and \emph{leader} servers to separate
functions (Section~\ref{sec:faaskeeper_design_writer},~\ref{sec:faaskeeper_design_distributor}).
Since functions are stateless, the entire state of a system must be stored in a replicated cloud storage (Section~\ref{sec:faaskeeper_design_storage}).
Serverless has new challenges: it does not have direct and ordered communication channels such as TCP connection, and we need to use ordered cloud queues and extend functions with logic to handle watch notifications and guarantee consistency.
Different types of storage for system and user data means we have separate data read and write paths, requiring extended system counters (Section~\ref{sec:faaskeeper_design_notification})
and additional ordering in \toolname{} client library (Section~\ref{sec:faaskeeper_design_client}).
Finally, the periodic heartbeat verification is mapped to a scheduled function 
(Section~\ref{sec:faaskeeper_design_heartbeat}).
A detailed discussion on how \toolname{} provides the same consistency
guarantees as Zookeeper can be found in Appendix~\ref{sec:appendix_faaskeeper}.

\DeclareRobustCommand{\numding}[1]{\normalsize\smash{\ding{\number\numexpr191+#1}}}
\DeclareRobustCommand{\lnnumding}[1]{
	\mbox{
		\raisebox{-0.2em}[0pt][0pt]{
			\normalsize\ding{\number\numexpr191+#1}
		}
	}
}

%\vspace{-0.5em}
\subsection{Follower}
\label{sec:faaskeeper_design_writer}
\toolname{} replaces ZooKeeper servers preparing update transactions with concurrently operating
\textbf{follower functions}.
%
%A cloud queue invokes functions (\circled{C2}), and the function processes requests of each client in a FIFO order (Alg.~\ref{alg:writer_function}).
A cloud queue invokes functions, and the function processes requests of each client in a FIFO order (Algorithm~\ref{alg:writer_function}).
The follower acquires a lock on the node (\numding{1}) to prevent concurrent updates,
verifies the correctness of the operation (\numding{2}), e.g.,
checking that a newly created node does not exist and the conditional update can be applied.
The validated and confirmed changes are propagated through a FIFO queue to the \emph{event \textbf{leader}} function (\numding{3}), ensuring that the changes are not reordered.
Finally, the new node version is secured in the system storage (\numding{4}) and extended with the current transaction's index.
This operation is combined with a lock release and applied conditionally, and no changes
are made if the lock expires.
At that point, the client request has been committed to the system (\nodeColor{Z1}{black}{ellipse}), and pushing to the queue before committing ensures that the leader will propagate the changes to the storage visible by users.
In some operations, the ZooKeeper model requires locking more than one node --- for example, creating a node also requires locking the parent node.
There, the commit creates a transaction from multiple atomic operations that will fail or succeed simultaneously.

Consecutive requests cannot be reordered, but the first stages of a request (\numding{1}, \numding{2})
can be executed while its predecessor is committed to the storage (\numding{3}, \numding{4}).
Thus, the follower function is a sequence of operations on the system storage
that can be \emph{pipelined}.

\noindent\textbf{Implementation.}
Each client session is assigned a queue to send new requests and invoke processing functions.
We select a cloud queue that fulfills the following requirements:
%(a) must invoke functions to process new data,
(a) invokes functions on messages,
(b) upholds FIFO order,
(c) allows limiting the concurrency of functions to a single instance,
(d) support batching of data items,
and (e) assigns monotonically increasing values to consecutive messages (\emph{txid}).
The requirements guarantee that requests are not reordered (\nodeColor{Z3}{black}{ellipse}),
while (d) ensures efficient processing of frequent invocations in a busy system.

\alglanguage{pseudocode}
\begin{figure}[t]
\vspace{-1em}
\begin{algorithm}[H]
\caption{A pseudocode of the new follower function.}
	\begin{algorithmic}
	\footnotesize
    \Function{Follower}{updates}
    \ForEachIn{client, node, op, args}{updates}
      \State lock, oldData = \textproc{Lock}(node) $\smash{\lnnumding{1}}$
      \If{\textbf{not} \textproc{IsValid}(op, args, oldData)} $\smash{\lnnumding{2}}$
        \State \textproc{Notify}(client, \textproc{FAILURE})
        \State \textbf{continue}
      \EndIf
      \State $txid$ = \textproc{LeaderPush}(client, lock, node, newData) $\smash{\lnnumding{3}}$
      \State \textproc{CommitUnlock}(lock, node, op, args, $txid$) $\smash{\lnnumding{4}}$
    \EndForEachIn
    \EndFunction
	\end{algorithmic}
\label{alg:writer_function}
\end{algorithm}
\vspace{-2.5em}
\end{figure}

\DeclareRobustCommand{\numdingDark}[1]{\normalsize\smash{\ding{\number\numexpr201+#1}}}
\DeclareRobustCommand{\lnnumdingDark}[1]{
	\mbox{
		\raisebox{-0.2em}[0pt][0pt]{
			\normalsize\ding{\number\numexpr201+#1}
		}
	}
}

\begin{figure}[t]
\vspace{-1em}
\begin{algorithm}[H]
	\caption{A pseudocode of the new leader function.}
	\begin{algorithmic}
	\footnotesize
    \Function{Leader}{state, updates}
    \ForEachP{region}
      \ForEachIn{client, lock, node, data, $txid$, followerID}{updates}
        \State nodeStatus = \textproc{GetNode}(node) $\smash{\lnnumdingDark{1}}$
        \If{nodeStatus.transactions[0] != $txid$}
            \If{\textbf{not} \textproc{TryCommit}(lock, node)} $\smash{\lnnumdingDark{2}}$
                \State \textproc{Notify}(client, \textproc{FAILURE})
                \State \textbf{continue}
            \EndIf
        \EndIf
        \State \textproc{DataUpdate}(region, data, $s'$, $epoch$) $\smash{\lnnumdingDark{3}}$
        \State w = \textproc{Watches}(node)
        \State \textproc{InvokeWatch}(region, w, \textproc{WatchCallback}) $\smash{\lnnumdingDark{4}}$
        \State $epoch[region] = epoch[region] + w$
        \State \textproc{Notify}(client, \textproc{SUCCESS})
        \State \textproc{PopTransaction}(node) $\smash{\lnnumdingDark{5}}$
      \EndForEachIn
    \EndForEachP
    \State \textproc{WaitAll}(\textproc{WatchCallback})
    \EndFunction
    \Function{WatchCallback}{epoch, region, w}
      \State $epoch[region] = epoch[region] - w$ $\smash{\lnnumdingDark{6}}$
    \EndFunction
	\end{algorithmic}
	\label{alg:distributor_function}
\end{algorithm}
%\vspace{-3em}
\end{figure}

%\vspace{-0.5em}
\subsection{Leader}
\label{sec:faaskeeper_design_distributor}
The \emph{leader} function (Algorithm~\ref{alg:distributor_function})
delivers updates to the cloud storage visible by users, similar to ZooKeeper's leader that distributes confirmed changes to servers handling read requests.
A FIFO queue between \emph{followers} and \emph{leader} is necessary to ensure that changes in user data stores are not reordered since concurrent updates could violate consistency (\nodeColor{Z3}{black}{ellipse}), and notifications must be delivered in order (\nodeColor{Z4}{black}{ellipse}).
Since a follower cannot push to the queue and commit the node atomically,
leader verifies that the node has been committed successfully (\numdingDark{2}).
In the case of the follower's failure or unlikely interleaving between both functions, 
the leader tries to commit nodes when possible to improve the system availability.
Otherwise, the update is rejected - the request has never been committed, and a failure of one follower function does not impact the system consistency.
Then, the data is replicated to user storage (\numdingDark{3}), and the leader sends watch notifications (\numdingDark{4}).
Committing changes to the user-visible storage must be serialized in ZooKeeper's consistency model,
and clients cannot observe newer data before receiving watch notifications (\nodeColor{Z4}{black}{ellipse}).
However, this process can be parallelized across cloud regions.
Once all steps are completed, the current transaction is removed from the node (\numdingDark{5}).
%to ensure that all notifications are always delivered
%
The per-node transaction index allows the cloud queue to retry the function invocation automatically
after a failure.

\noindent\textbf{Implementation.}
When committing data to cloud storage, we attempt to
update only changed data to avoid unnecessary costs and network traffic.
While early visions of object storage assumed that write operations can access arbitrary offsets in an object~\cite{1612479,1194853}, these are not widely available in modern clouds.
%While this is common for NoSQL databsases, the update operation of S3
%requires the complete replacement of data.
%
Thus, even if a change involves only part of node’s metadata, the leader
function needs to download entire node first %from storage
to conduct the update operation.

%\vspace{-1em}
\subsection{Storage}
\label{sec:faaskeeper_design_storage}
ZooKeeper achieves high availability with multiple replicas of the dataset.
We achieve the same goal by using automatically replicated and scalable cloud storage,
which helps us to simplify the control plane of our system.
We distinguish two types of storage in \toolname{}: \textbf{system storage}
used by followers and leader to coordinate and modify the system state,
and \textbf{user storage} optimized to handle read requests from \toolname{} clients in a scalable and cost-efficient manner.
\textbf{System storage} contains the current timestamp, all active sessions, and the list of all data nodes to allow locking by follower functions.
\textbf{Data storage} is indexed by node paths, and each item corresponds to a single ZooKeeper node,
with user data, modification timestamps, and a list of node children.

When selecting an instance of cloud storage for \toolname{}, we consider not only the cost and performance but also the technical capabilities of the services.
Eventually consistent reads neither guarantee read--your--write
consistency~\cite{vogels2009eventually}, nor consider a dependency between
different writes, breaking ZooKeeper guarantees
(Linearized Writes \nodeColor{Z2}{black}{ellipse},
Single System Image \nodeColor{Z3}{black}{ellipse}).
Therefore, we must use cloud storage that supports strongly consistent reads.

\noindent\textbf{Synchronization Primitives}
%
%are implemented with \emph{update expressions} of DynamoDB system storage~\cite{awsDBExpr}.
are implemented in system storage and require that each update to a single item is atomic.
Atomic counters are implemented as a single number, and an update adds a numerical constant to the current value.
Atomic lists are represented as a list of numbers with an update that adds and removes
elements from the list.
Finally, the timed lock uses conditional updates to verify that each locking and unlocking operation
does not invalidate any existing locks.
A lock is stored in the node as a timestamp, allowing other functions to override an expired lock and prevent deadlocks caused by a failure in a function.
Each operation requires a single write to a single item.

\noindent\textbf{Timestamps} provide an order over system transactions.
To guarantee the consistency of updates,
we need to define a total ordering of modifications in the system.
on the "happened before" relation~\cite{10.1145/359545.359563}.
The system \textbf{\emph{state} counter} $txid$ is an integer that represents 
the \textbf{timestamp} of each change in \toolname{}, similar to the \emph{zxid} state 
counter in ZooKeeper and provides total order over the system.
Each transaction modifies the state counter atomically,
providing a total ordering of all processed modification requests.
The \textbf{\emph{epoch} counter} is specific to \toolname{} and contains watch notifications
pending while the transaction represented by the state counter was in progress.
Counters are implemented using the atomic counters and lists 
(Section~\ref{sec:faaskeeper_serverless_components}).

%\vspace{-1em}
\subsection{Watch Notifications}
\label{sec:faaskeeper_design_notification}
When a ZooKeeper client changes a node that has watches attached to it,
the system sends a notification to watch owners, who must not see new data before receiving a notification.
In serverless, the path of reads and writes are different, a significant departure from ZooKeeper, where all reads and writes are processed by the same entity, and the underlying TCP connection guarantees order.
Instead, we use the additional region-wide \textbf{epoch counters} 
to provide an ordering between notifications and changes applied to the system.
Each watch is assigned a unique identifier, and multiple clients can be assigned to a single watch instance.
When a node is updated, the leader attaches to it the epoch counter containing the identifiers of all watch notifications still being delivered while the update was taking place.
Once the client library finds the counter in a read node, it checks the epoch counter for any of the watches registered by this client.
In such a case, the read operation must be stalled until the pending notification is delivered,
preventing the client from seeing updated data before observing all preceding notifications.
%

%\vspace{-1em}
\subsection{Client}
\label{sec:faaskeeper_design_client}
\toolname{} implements the same standard read and write operations as ZooKeeper and offers clients an API similar to ZooKeeper.
\emph{Read} operations are served with a direct access to the cloud storage.% (\circled{C1}).
\emph{Write} operations are sent to \emph{follower} functions through a cloud queue.
% (\circled{C2}).
%
Eliminating the server from the data access path provides lower operating costs,
but puts on the client the responsibility of ordering results with watch notifications.
Thus, we replace the event coordination on ZooKeeper servers with a lightweight queue on the client: a read following a write cannot return before its predecessor.

\noindent\textbf{Implementation.}
Each client runs three background threads to send requests, manage
incoming responses, and order results.
Epoch counters ensure the ordering of writes and notifications, and queues replace the ordered TCP communication of ZooKeeper.

%\vspace{-1em}
\subsection{Heartbeat}
\label{sec:faaskeeper_design_heartbeat}
In addition to ordering guarantees, sessions play another significant role in ZooKeeper:
their status defines the lifetime of ephemeral nodes, which are automatically deleted upon the
closure of their owner's session.
We replace the heartbeat messages with a \emph{scheduled \textbf{heartbeat}} function
to prune inactive sessions and notify clients that the system is online.

\textbf{Implementation.}
The cloud system periodically invokes the function which sends in parallel heartbeat messages
to clients that own ephemeral nodes.
If a client does not respond before a timeout,
the function begins an eviction process for the session by placing a
deregistration request in the processing queue.
The function is parameterized with the \emph{heartbeat frequency} parameter $H_{fr}$.

\subsection{Summary}

To finalize the serverless redesign of an IaaS service, we
need to incorporate an elastic scaling model and ensure cloud portability.

\noindent \textbf{Elastic resource allocation.}
To accommodate the temporal and spatial irregularity of workloads~\cite{4362193},
\toolname{} attempts to scale the resource allocation linearly with the demand.
In the case of a \emph{shutdown}, the user should pay only for keeping
the data in the cloud.
Therefore, we use the pay-as-you-go billing scheme of the storage and
queue services, dependent only on the number of operations performed
and not on the resources provisioned.

\noindent \textbf{Cloud agnosticity.}
Vendor lock-in~\cite{6415917} is a serious limitation in serverless~\cite{9218932,10.1145/3472883.3487002},
and dependency on queueing and storage services is of particular concern~\cite{10.1145/3106237.3117767}.
FaaS applications implemented in a specific cloud often include provider-specific
solutions, requiring a redesign and reevaluation of the architecture
when porting to another cloud.
In the \emph{cloud-agnostic} design of \toolname{}, we define only the requirements for each service used and introduce new
abstractions such as synchronization primitives to encapsulate cloud-specific solutions.
We specify expectations on serverless services at the level of semantics and guarantees.
This limits our dependency
%to vendor lock-in
to the implementation layer
and allows moving between providers without a major system overhaul~\cite{10.1007/978-3-642-24755-2_6}.

%-------------------------------------------------------------------------------

%-------------------------------------------------------------------------------
\section{From FaaSKeeper Design To Cloud}

\label{sec:faaskeeper-optimizations}

\begin{table}[t!]
    \centering
    \begin{adjustbox}{max width=\linewidth}
    %\normalsize
    \footnotesize
    \begin{tabular}{p{0.8cm}c|C{1.4cm}|C{1.6cm}|C{1.5cm}|C{1cm}}
      %\toprule
                    & \textbf{Requirements} & \textbf{AWS} & \textbf{Azure} & \textbf{Google}
                   & \textbf{Other}\\

\hline
\multirow{3}{*}{\textbf{Function}} & Free        & \cmark  & \cmark    & \cmark  & \multirow{3}{*}{---} \\
                   & Event       & \cmark  & \cmark    & \cmark\\
                  & Scheduled   & \cmark  & \cmark    & \cmark\\
\hline
\textbf{User}     &                           & S3          & Blob Storage  & Storage   & Redis\\
\textbf{Store}    & Consistency        & \cmark  & \cmark    & \cmark &  \cmark \mmark\\
                  & Throughput      & \cmark  & \cmark    & \cmark &  \cmark \mmark\\
\hline
                      &                       & DynamoDB  & CosmosDB  & Datastore   & Redis\\
  \textbf{System}     & Reliability           & \cmark  & \cmark    & \cmark &  \xmark  \\
  \textbf{Store}      & Consistency    & \cmark  & \cmark    & \cmark &  \cmark \mmark\\
      & \multirow{2}{*}{\minitab[c]{Concurrency\\Primitives}}
      %& \multirow{2}{*}{\minitab[c]{Conditional\\Updates}}
      & Conditional
      & Optimistic
      & Transactions
      & Lua\\
      %& \multirow{2}{*}{Conditional Updates}
      %& \multirow{2}{*}{\minitab[c]{Optimistic\\Locking}}
      %& \multirow{2}{*}{\minitab[c]{Read-Write\\Transactions}}
      %& \multirow{2}{*}{\minitab[c]{Server-Side\\Transactions}}\\
       & &   Updates   & Locking && Scripts\\
\hline
\multirow{3}{*}{\textbf{Queue}}         &               & SQS       & Service Bus & Pub/Sub   & \multirow{3}{*}{---} \\
                                        & FIFO          & \cmark    & \cmark    & \cmark      &\\
                                        & Serverless    & \cmark    & \cmark \mmark & \cmark      &\\
    \end{tabular}
  \end{adjustbox}
  \caption{\textbf{Mapping \toolname{} design to cloud and user-managed services. \mmark\xspace indicates additional constraints.}}
  %\vspace{-3em}
  \label{tab:cloud_mapping}
\end{table}

In the previous section, we mapped the ZooKeeper components to a cloud-agnostic design with separate services.
Now, we map \toolname{} functions, queue, storage, and
synchronization primitives to actual cloud services (Table~\ref{tab:cloud_mapping}),
tailoring resource requirements to each component and enabling serverless scalability.
Following the multi-step design guideline,
we disaggregate computing (Section~\ref{sec:faaskeeper_disaggregating}),
incorporate different types of cloud storage (Section~\ref{sec:faaskeeper_cloud_data_storage}),
and find the most optimal ways of exchanging data in the system (Section~\ref{sec:faaskeeper_communicating}).
Compute tasks can now be fully serverless, and the system requires no resource provisioning
while providing compatible interface to ZooKeeper clients 
(Section~\ref{sec:faaskeeper_compatibility}).
We scale \toolname{} up by adding more concurrent \emph{follower} functions and placing data in different storage keys to benefit from sharding and cloud scalability.

\noindent\textbf{Implementation}.
We select the AWS cloud and translate design concepts to cloud systems: system storage with
DynamoDB tables, synchronization primitives to DynamoDB \emph{update expressions}~\cite{awsDBExpr}, user data storage to S3 buckets, and FIFO queues to the SQS.
We use the AWS SQS with batched Lambda invocations~\cite{awsSQSLambda}
as it performs better than DynamoDB Streams (Section~\ref{sec:evaluation_queues}).
We implement the four \toolname{} functions in 1,350 lines of Python code in
AWS Lambda.
Furthermore, we provide a client library with 1,400 lines of Python code with the relevant methods of
the API specified by ZooKeeper~\cite{10.5555/1855840.1855851}.
Each component has a corresponding alternative in other cloud systems that provides the same semantics
and guarantees, and storage
can be improved with in-memory caches.% such as Redis.

%\vspace{-1.25em}
\subsection{Disaggregating}
\label{sec:faaskeeper_disaggregating}
Although ZooKeeper servers manage connections and ordering,
their primary responsibility is to provide low-latency data access
that can be replaced with cloud storage.
In a coordination system designed for high read-to-write ratios,
more resources should be allocated for data endpoints rather than servers handling computing tasks.
In \toolname{}, we removed the need for separate \emph{reader} function - clients
access cloud storage directly, saving time and money.
Both key-value and object storage implement replication, with DynamoDB using three-way replication of each partition and S3 guaranteeing 11 9's of durability for each object.
For that reason, we do not have to implement any additional replication.

\begin{figure}
	\centering
  \begin{subfigure}{\linewidth}
    \includegraphics[width=\textwidth]{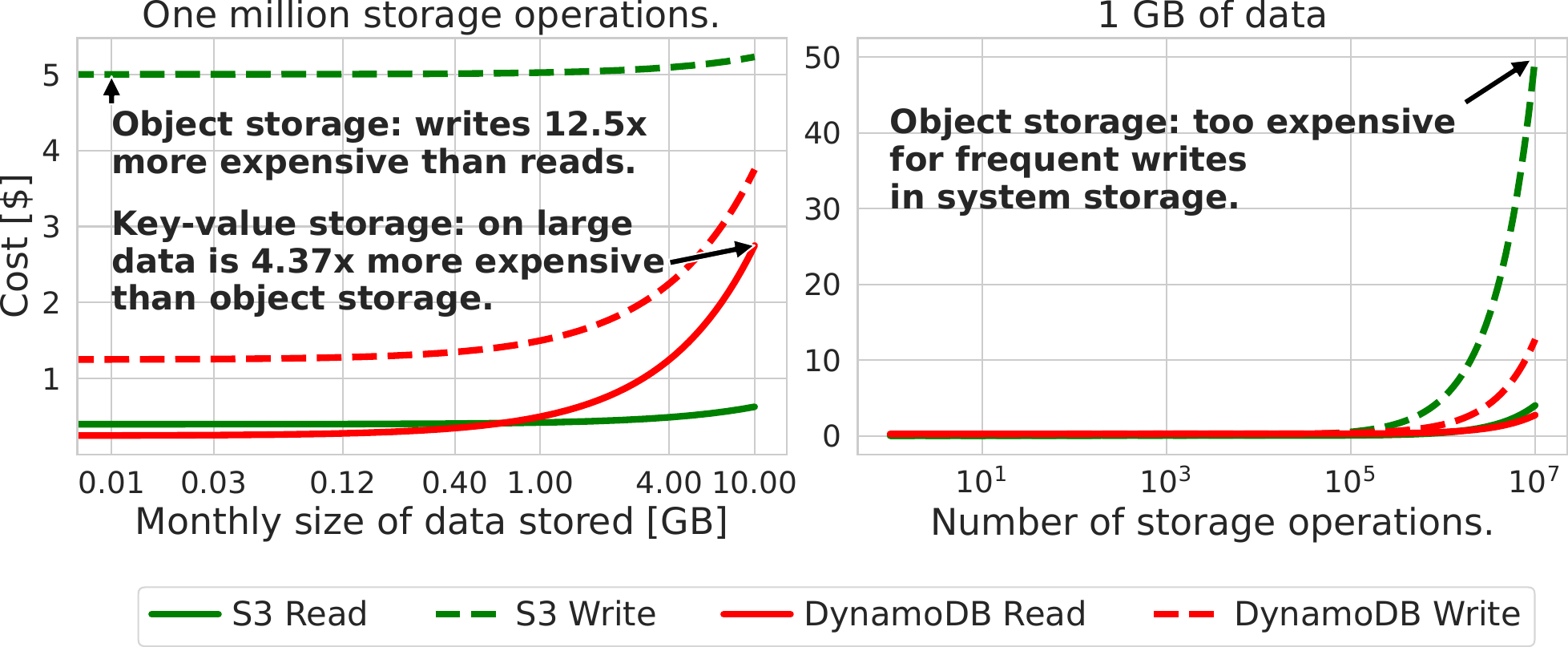}
    \vspace{-1em}
    \caption{Cost of storage services for varying data size and 1 kB operations.}
    \label{fig:microbenchmark_storage_cost}
  \end{subfigure}
	\\
  \begin{subfigure}{\linewidth}
    \includegraphics[width=\textwidth]{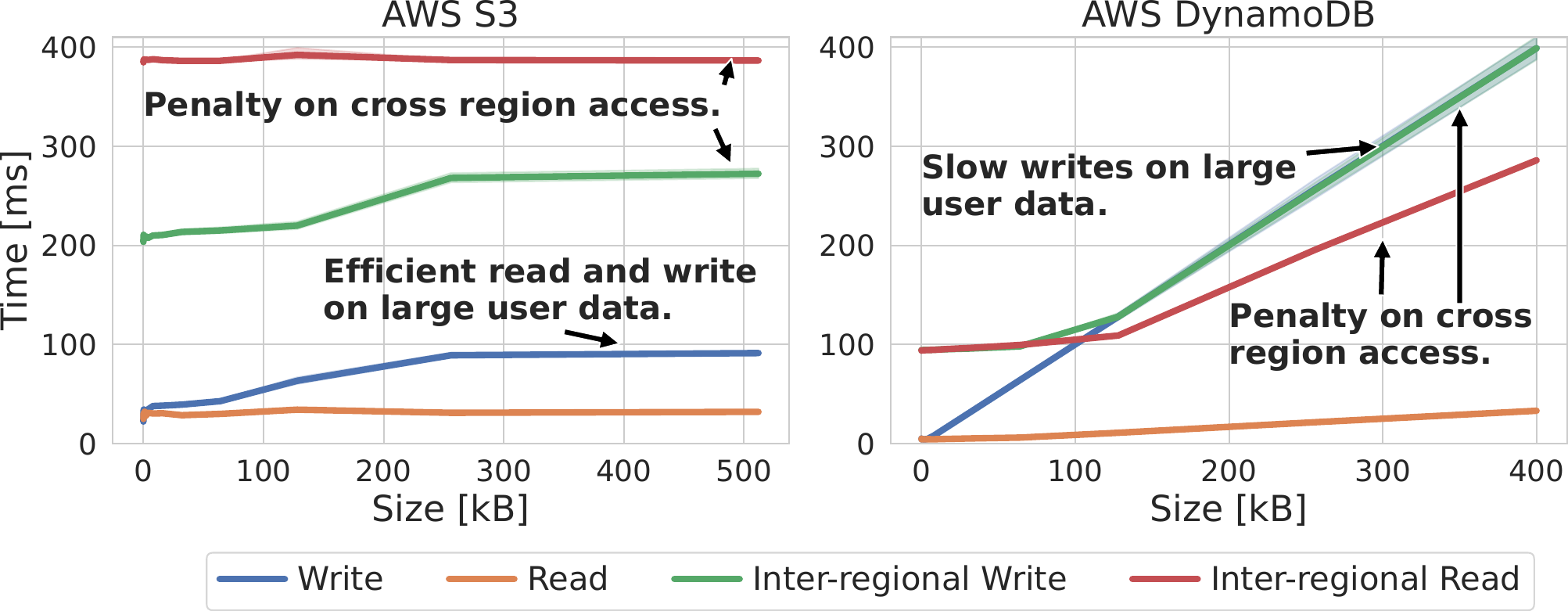}
    \vspace{-1em}
    \caption{Latency of read and write operations in AWS storage services.}
    \label{fig:microbenchmark_storage_perf}
  \end{subfigure}
  \vspace{-1em}
  \caption{\textbf{Cost and performance of storage in the AWS cloud. Python benchmarks executed on an EC2 instance.}}
  \label{fig:microbenchmark_storage}
\end{figure}

\noindent \textbf{User data locality.}
Cloud applications balance resource allocation across
geographical regions to support changing workloads~\cite{10.1145/2987550.2987561,10.1007/978-3-642-13119-6_2}.
In addition, they aim to minimize the distance between the service and its users,
as the cross-region
%network
transmission
adds major performance and cost overheads (Figure~\ref{fig:microbenchmark_storage_perf}).
While ZooKeeper requires migrating a virtual machine
across regions~\cite{10.1145/2987550.2987561},
\toolname{} can serve data from endpoints local to the user.
Clients connect to the closest %auto-scaling
storage in their region, minimizing access latency.

\noindent\textbf{Decoupling Watch Delivery}
In \toolname{}, we moved the delivery of watch notifications into a separate \emph{free function \textbf{watch}}.
Since hundreds of clients can register a single watch, using a serverless function allows us to adjust
resource allocation to the workload.
A standard writing pipeline includes only querying watch information in the system storage, adding insignificant cost and overhead.

\subsection{Mapping Storage}
\label{sec:faaskeeper_cloud_data_storage}
With storage and computing decoupled, we can map them to services
that fit best their access patterns and computational requirements.
Storage should distinguish between user data and the system data needed to control ZooKeeper:
locality and cost requirements are different, and storage solutions have varied costs and latencies
-- especially when different sizes are considered (Figure~\ref{fig:microbenchmark_storage}).

\noindent\textbf{Efficient reading of user data.}
ZooKeeper is optimized for high read performance.
Thus, we must use storage with strongly consistent, cheap, and fast read operations.
The cost-performance analysis reveals that object storage is
more efficient than key-value storage (Figure~\ref{fig:microbenchmark_storage_cost}).
Storing large user data is 4.37x cheaper,
and updating nodes scales much better with their size.
Furthermore, read operations are billed per access and per \SI{4}{\kilo\byte} read in object and key-value
storage, respectively, making the latter more expensive for large data in user nodes by even an order of magnitude.

Furthermore, we optimize ZooKeeper's \code{get\_children} operation by storing the children list in the
metadata of each node.
This update does not add costs, as adding and removing nodes requires locking and updating the parent, and we avoid the expensive scan.% operation.% to find all children.

\noindent \textbf{Efficient modifications of control data.}
The system state includes frequently modified watches,
client and node status, and synchronized timestamps.
\toolname{} must use atomic operations and locks to support concurrent updates.
We use the key-value store as the object store is limited by
expensive
writes to small items (Figure~\ref{fig:microbenchmark_storage_cost})
and lack of synchronization primitives.

\noindent \textbf{Hybrid storage}
While DynamoDB is cheaper for small nodes and faster (Section~\ref{sec:evaluation_reads}),
the costs explode for large user data, restricting us to object storage.
However, even though nodes store up to 1 MB of data, the dominant use case of ZooKeeper is
to store small configuration objects.
Thus, we optimize for the common case and place in DynamoDB all nodes up to 4 kB,
and split node metadata and user data between DynamoDB and S3 for larger nodes.
The client library begins by reading data from key-value storage,
and only the infrequent large nodes incur the performance and cost penalty of a second storage request.
This allows us to improve read latency by over 50\% (Section~\ref{sec:evaluation_reads})
and decrease costs by 37.5\%.

\subsection{Communicating Functions}
\label{sec:faaskeeper_communicating}
\toolname{} functions scale automatically with workload and emulate the TCP connection between the client and ZooKeeper servers. 

\noindent \textbf{Vertical scaling.}
ZooKeeper improves throughput by \emph{pipelining} client requests over a single TCP
connection to the server.
Requests are sent before previous operations finish, and the implementation ensures that
operations from a single session are not reordered in the pipeline.
However, serverless functions are designed for fine--grained invocations.
Thus, \toolname{} employs cloud queues to batch invocations and continuously feeds the processing
pipeline.
However, cloud queues invoke functions in batches, preventing continuous streaming of new requests to the pipeline.

\noindent \textbf{Horizontal scaling.}
ZooKeeper achieves high read scalability with more servers, but write scalability is limited
by design with a single leader.
Prior attempts to increase write performance focused on
partitioning the ZooKeeper data tree~\cite{8023134,6983381}.
Instead, \toolname{} delegates requests from different client sessions to concurrently
operating functions.
While write requests of a single session are serialized, we exploit the parallelism
of operations from different users.
To determine global ordering and handle concurrent modifications to the same data node,
\toolname{} uses synchronization primitives on the storage (Section~\ref{sec:faaskeeper_serverless_components}).
Thus, the scalability of read and write operations is bounded by storage throughput.

\subsection{Compatibility with ZooKeeper}
\label{sec:faaskeeper_compatibility}
Our implementation is standalone and does not reuse the server-centric ZooKeeper codebase since Java
functions are by large cold startup overheads~\cite{8605773,10.5555/3277355.3277369}.
We offer a compatible interface for existing applications by modeling our
API after kazoo~\cite{kazoo}, a Python client for ZooKeeper.
While \toolname{} aims to provide consistency model and interface compatible with ZooKeeper,
we make minor adjustments due to the limitations of cloud services and the serverless model.
While large ZooKeeper nodes are uncommon and impractical, we can support the 1MB node
in cloud object storage.
The size restrictions of 400 and 256 kB in DynamoDB, respectively,
limit the maximum data sent by users.
This can be avoided by splitting larger nodes and using temporary S3 objects.
Furthermore, Zookeeper clients can define node permissions with access control lists (ACLs).
In \toolname{}, functions implement write permissions thanks to the
protection boundary between caller and callee, and read permissions can
be enforced with cloud storage ACLs.

%\vspace{-1em}
\subsection{Cloud Portability}
To validate that \toolname{} design is cloud-agnostic and not locked to a single provider,
we ported it to the Google Cloud Platform.
We replace cloud services as specified in Table~\ref{tab:cloud_mapping},
and achieve the same semantics of a serverless service with pay-as-you-go-billing.
The majority of the implementation effort was in adapting to new APIs and adding synchronization primitives as transactions~\cite{googleDBTransactions}, with changes in the system library (600 LoC), client code (200 LoC), and configuration (150 LoC).
Google Cloud has size limits of 10 MB and 1 MB on queue and key-value storage operations, respectively,
simplifying the implementation of large ZooKeeper nodes.

However, both platforms have different pricing models that affect the optimizations.
While object storage costs the same, operations pricing on the key-value Datastore is independent of the item size.
Compared to AWS DynamoDB, Datastore is 2.4x and 1.44x more expensive on read and write operations of up to 1 KB, respectively.
While this simplifies the system design as we no longer need special treatment for large nodes,
this is not the common case for ZooKeeper.
On the other hand, the Pub/Sub queue charges clients based on the amount of data sent and received,
but not less than 1 KB per message.
At \$40 per terabyte of data, the queue is 6.7x cheaper for small messages than AWS SQS, 
which charges \$0.5 per one million messages.

%-------------------------------------------------------------------------------

%-------------------------------------------------------------------------------
\section{Evaluation}

\label{sec:faaskeeper_evaluation}

We begin with analyzing ZooKeeper utilization (Section~\ref{sec:zookeeper_utilization})
and benchmarking serverless components necessary to build a serverless service (Section~\ref{sec:evaluation_components}).
Then, we evaluate the performance-cost trade-offs of \toolname{} in relation to ZooKeeper (Section~\ref{sec:evaluation_fk_zk}).
We answer the following questions:

\begin{itemize}[noitemsep,topsep=0pt]
  \itemsep0em
  \item[\S~\ref{sec:zookeeper_utilization}] How frequently is ZooKeeper used in practice?
  \item[\S~\ref{sec:evaluation_primitives}] Are synchronization primitives efficient?
  \item[\S~\ref{sec:evaluation_queues}] Do serverless queues provide cheap and fast invocations?
  \item[\S~\ref{sec:evaluation_reads}] How fast are cloud-native read requests in \toolname{}?
  \item[\S~\ref{sec:evaluation_writes}] How expensive is the processing of write requests?
  \item[\S~\ref{sec:evaluation_heartbeat}] What are the cost savings in service monitoring?
  \item[\S~\ref{sec:faaskeeper_properties}] What is the cost break-even point for \toolname{}?
\end{itemize}

\emph{Evaluation Platform}
The deployment in the AWS region \sloppy\code{us-east-1} consists of four functions, SQS queues, and \code{DynamoDB} tables storing system state, user list, and watches.
Functions are allocated with 2048 MB of memory if not specified otherwise.
Additionally, we use a \code{DynamoDB} table or an S3 bucket for user data storage.
Benchmarks use Python 3.8.10, and we run microbenchmarks and
\toolname{} clients from a \code{t3.medium} virtual machine with \code{Ubuntu 20.04}
in the same cloud region.
We also deploy \toolname{} in the GCP region \sloppy\code{us-central1}.
Benchmarks use Python 3.8.10, and we run benchmarking clients from a \code{e2-medium} virtual machine with \code{Ubuntu 20.04} in the same cloud region.
ZooKeeper 3.7.0 is deployed on three VMs running \code{Ubuntu 20.04},
using \code{t3.small} and \code{e2-small} machines on AWS and GCP, respectively.

\begin{figure}
    \centering
    \includegraphics[width=\linewidth]{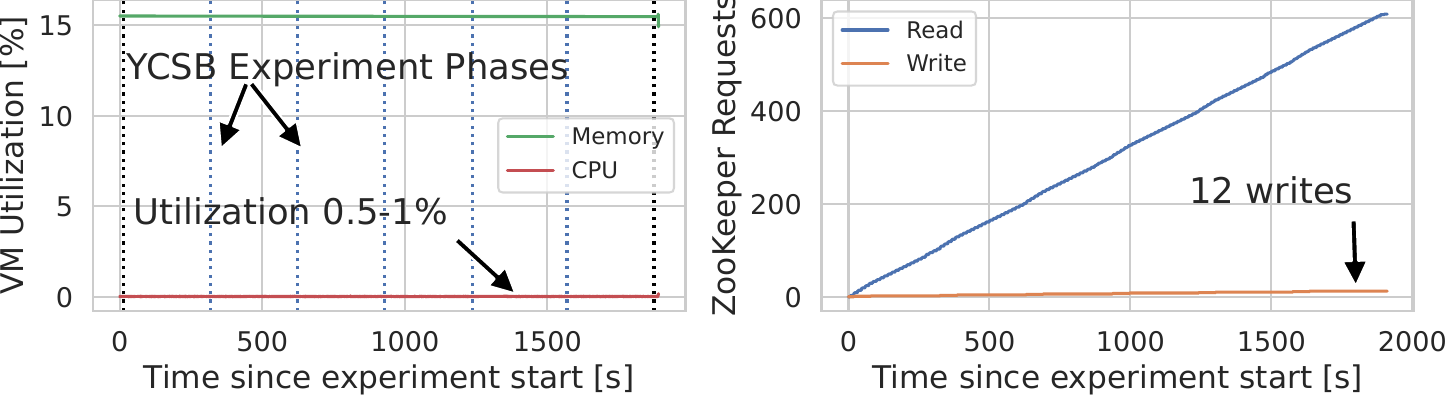}
    \vspace{-1.5em}
    \caption{ZooKeeper utilization in HBase running YCSB.}
    \label{fig:zookeeper_utilization}
\end{figure}

\subsection{ZooKeeper}
\label{sec:zookeeper_utilization}
To understand how ZooKeeper is used in practice, we profile its utilization in Apache HBase.
We deploy HBase 2.5.6 with Hadoop 3.3.2 on four \code{t3.2xlarge} machines, with one holding HDFS NameNode
and HBase HMaster, and others serving data, and ZooKeeper 3.7.2 on three \code{t3.medium} machines.
From the benchmarking virtual machine \code{t3.2xlarge}, we execute the standard workloads from YCSB~\cite{10.1145/1807128.1807152}, each running for five minutes, and present results
in Figure~\ref{fig:zookeeper_utilization}.
The HBase service can handle thousands of requests while using ZooKeeper only to control the state of the cluster, 
With less than a thousand requests in over half an hour, replacing persistent ZooKeeper with a serverless system is a significant optimization opportunity.

Furthermore, we analyzed the size of ZooKeeper nodes after the experiment ended.
HBase created 29 nodes, with a median and mean data size of 0 and 46 bytes, respectively.
The largest node had 320 bytes of data and corresponded to each \emph{RegionServer}.

\begin{figure}[t!]

  \begin{subfigure}[b]{\linewidth}
    \centering
    \begin{adjustbox}{max width=\linewidth}
    \footnotesize
    \begin{tabular}{l|r|rrrrr}
      %\toprule
      Primitive & Size & \textbf{Min} & \textbf{p50} & \textbf{p95} & \textbf{p99} & \textbf{Max}\\
      \midrule
      \multirow{2}{*}{\parbox{2cm}{Regular\\DynamoDB \textbf{write}}}
                    &1 kB  &  3.95  &  4.35  &  4.79  &  6.33  &  60.26\\
                   & 64 kB  &  6.54  &  66.31  &  70.28  &  77.23  &  121.64\\
      \midrule
      \multirow{2}{*}{\parbox{2cm}{ Timed lock\\ \textbf{acquire}}}
                  & 1 kB &  6.13  &  6.8  &  8.13  &  14.14  &  65.32\\
                  & 64 kB &  7.82  &  67.16  &  72.71  &  90.56  &  177.02\\
      \midrule
      \multirow{2}{*}{\parbox{2cm}{ Timed lock\\ \textbf{release}}}
                    & 1 kB &  6.03  &  6.62  &  7.94  &  12.52  &  78.44\\
                    & 64 kB &  6.38  &  65.2  &  70.33  &  92.15  &  222.64\\
      \midrule
      \multirow{1}{*}{\parbox{2cm}{Atomic counter}}  & 8 &  4.88  &  5.59  &  7.01  &  11.69  &  62.4\\
      \midrule
      \multirow{2}{*}{\parbox{2cm}{Atomic list\\ \textbf{append} } }   & 1  &  5.14  &  5.89  &  8.0  &  10.71  &  21.12\\
                                                & 1024  &  16.72  &  76.01  &  184.02  &  187.47  &  249.23\\
      %\bottomrule
    \end{tabular}
    \end{adjustbox}
    \caption{Latency of synchronization primitives for varying item size (lock) and list append length (atomic list).}
    \label{tab:synchronization_primitives}
  \end{subfigure}

	\hfill
  \begin{subfigure}{\linewidth}
    \centering
    \includegraphics[width=\linewidth]{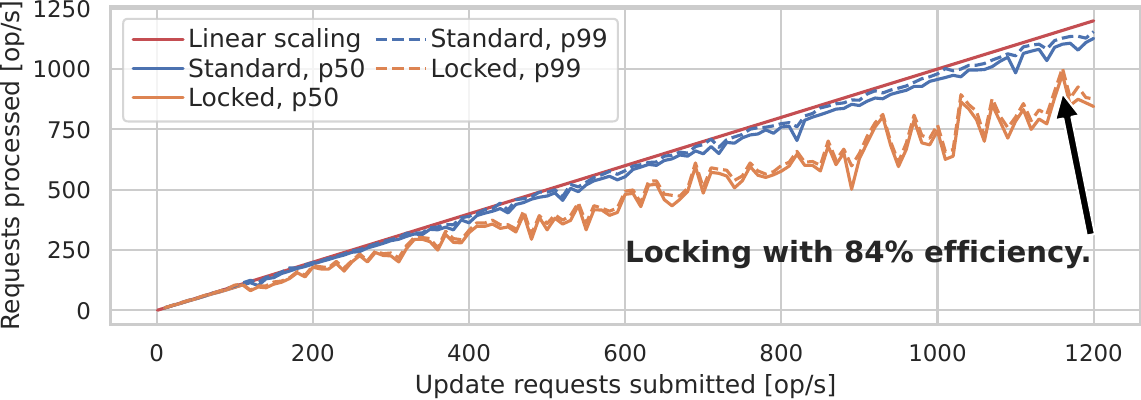}
    \caption{
      Throughput of standard and locked DynamoDB updates.
    }
    \label{fig:synchronization_primitives}
  \end{subfigure}
  \caption{Synchronization primitives on AWS DynamoDB.}
\end{figure}

\subsection{Serverless Components}
\label{sec:evaluation_components}

Now we evaluate the latency and throughput of components necessary to build a serverless service (Section~\ref{sec:faaskeeper_serverless_components}).

\subsubsection{Synchronization Primitives}
\label{sec:evaluation_primitives}
The serverless synchronization primitives bring concurrent and safe updates to \toolname{}.
Primitives are implemented with conditional update expressions of DynamoDB~\cite{awsDBExpr},
and we evaluate the overheads and scalability of this datastore system.

\noindent\textbf{Latency.}
We evaluate each operation by performing 1000 repetitions 
on warmed-up data and present results in Table~\ref{tab:synchronization_primitives}.
Each \textbf{timed lock} operation requires adding 8 bytes to the timestamp.
However, the operation time increases significantly with the item size, even though
large data attributes are neither read nor written in this operation.
This conditional and custom update adds 2.5 ms to the median time of a regular DynamoDB write,
and large outliers further degrade the performance.
This result further proves the need to disaggregate the frequently modified
system storage from the user data store, where items can store hundreds of kilobytes of data.
Then, we evaluate the \textbf{atomic counter} and \textbf{atomic list} expansion
by adding a varying number of items of 1 kB size.
This allows users to add new watches in storage with a single operation.

\noindent\textbf{Throughput.}
Timed locks allow \toolname{} to conduct independent updates concurrently.
We evaluate a pair of regular reads and writes, compare  ing them against our locks with a safe parallelization.
We measure the median throughput over a range of five seconds and vary the
workload, as well as the number of processes sending requests.
We use the \code{c4.2xlarge} VM as a client to support this multiprocessing
benchmark (Figure~\ref{fig:synchronization_primitives}).
Even though locks increase the latency of the update operation, the locked version
still achieves up to 84\% efficiency when handling over 100 requests per second
from ten clients concurrently.
This result agrees with previous findings that DynamoDB scales up to thousands
of transactions per second~\cite{10.1145/3342195.3387535}, and the
throughput of operations on DynamoDB is limited by Lambda's parallelism
and not by storage scalability~\cite{258880}.
\begin{summaryBox}{boxOliveGreen}
Our synchronization primitives introduce a few milliseconds of overhead
per operation and allow for parallel \toolname{} writes of
up to 1200 requests per second.
\end{summaryBox}

%\vspace{-1em}
\subsubsection{Serverless Queues}
\label{sec:evaluation_queues}
Queues improve the writing process by batching requests and 
are necessary to provide ordering (Section~\ref{sec:faaskeeper_design_distributor})
AWS offers two cloud-native queues with pay-as-you-go billing and function
invocation on new messages: SQS and DynamoDB Streams.
For \toolname{}, we select a queue that adds the minimal invocation
overhead and allows to achieve sufficient throughput.
%
% not ideal bu LatEx inserts a pause here.
%\linebreak
%\emph{Queue.}
%
For SQS~\cite{awsSQS}, we enable the FIFO property that comes with the restriction
of a maximum batch size of 10.
We compare against the standard version to estimate the potential overhead
of small batch sizes.
For DynamoDB streams, we configure database sharding to guarantee that
all items in a table are processed in order~\cite{awsDynamoLambda}.
We restrict the function's concurrency to permit only one instance at a time.

\begin{figure}[t]
\begin{subfigure}[b]{\linewidth}
  %\begin{table}
    \centering
    \begin{adjustbox}{max width=\linewidth}
    \footnotesize
    \begin{tabular}{lcc|cc|cc|cc}
      %\toprule
      %Primitive & \multicolumn{4}{c}{SQS} \\
       & \multicolumn{2}{c}{\textbf{Direct}} & \multicolumn{2}{c}{\textbf{SQS}} & \multicolumn{2}{c}{\textbf{SQS FIFO}} & \multicolumn{2}{c}{\textbf{DynamoDB Stream}} \\
         %\cline{2-4} \cline{5-7} \cline{8-10}
         & 64B & 64 kB & 64B & 64 kB & 64B & 64 kB & 64B & 64 kB \\
         \cmidrule{2-3} \cmidrule{4-5} \cmidrule{6-7} \cmidrule{8-9}
  \multicolumn{1}{c}{\textbf{p50}}    &       39.0    &       48.69   &       39.83   &       51.68   &       24.22   &       34.47   &       242.65  &       237.75\\
  \multicolumn{1}{c}{\textbf{p95}}    &       73.92   &       83.36   &       78.29   &       138.94  &       84.29   &       44.7    &       270.63  &       262.61\\
  \multicolumn{1}{c}{\textbf{p99}}    &       124.01  &       117.25  &       125.24  &       184.91  &       162.42  &       55.26   &       417.21  &       464.52\\
  \multicolumn{1}{c}{\textbf{Max}}    &       210.11  &       129.15  &       295.01  &       211.55  &       172.48  &       109.61  &       749.16  &       610.96\\

    \end{tabular}
  \end{adjustbox}
  \caption{End-to-end latency of FaaS invocation on AWS with a TCP reply.}
  \label{tab:queue_latency}
  %\end{table}
\end{subfigure}

	\hfill
  \begin{subfigure}{\linewidth}
    \centering
    \includegraphics[width=\linewidth]{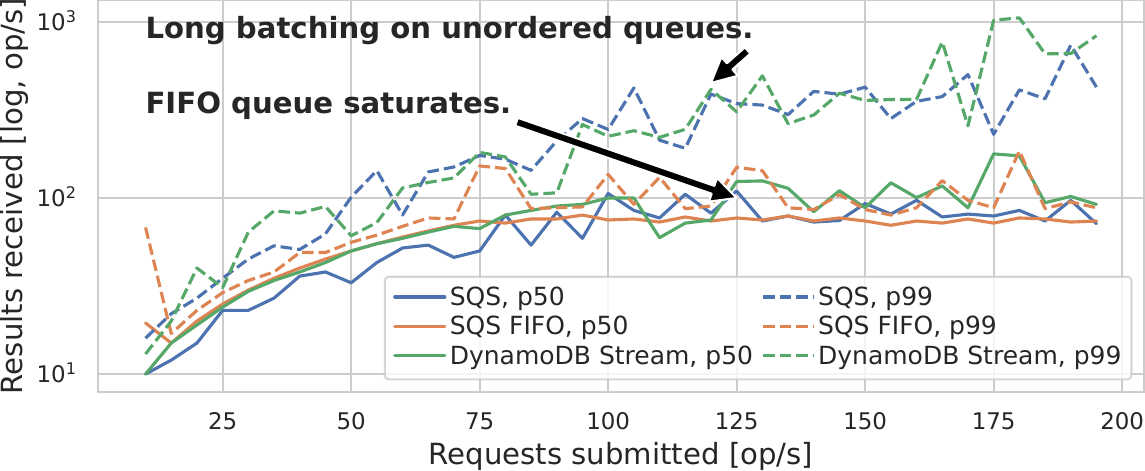}
    \caption{
      Throughput of function invocations on queues with 64B payload.
    }
    \label{fig:faaskeeper_performance_read}
  \end{subfigure}
\begin{subfigure}[b]{\linewidth}
  %\begin{table}
    \centering
    \begin{adjustbox}{max width=\linewidth}
    \footnotesize
    \begin{tabular}{lcc|cc|cc}
      %\toprule
      %Primitive & \multicolumn{4}{c}{SQS} \\
       & \multicolumn{2}{c}{\textbf{Direct}} & \multicolumn{2}{c}{\textbf{PubSub}} & \multicolumn{2}{c}{\textbf{PubSub FIFO}} \\
         %\cline{2-4} \cline{5-7} \cline{8-10}
         & 64B & 64 kB & 64B & 64 kB & 64B & 64 kB\\
         \cmidrule{2-3} \cmidrule{4-5} \cmidrule{6-7}
  \multicolumn{1}{c}{\textbf{p50}}    &       83.29   &       85.29   &       38.04   &       29.23   &       201.22  &       206.62\\
  \multicolumn{1}{c}{\textbf{p95}}    &       94.63   &       95.61   &       95.77   &       39.46   &       234.8   &       250.46\\
  \multicolumn{1}{c}{\textbf{p99}}    &       112.74  &       97.49   &       114.43  &       46.32   &       581.19  &       263.0\\
  \multicolumn{1}{c}{\textbf{Max}}    &       1115.14 &       112.73  &       643.96  &       57.66   &       588.95  &       280.84\\

    \end{tabular}
  \end{adjustbox}
  \caption{End-to-end latency of FaaS invocation on GCP with a TCP reply.}
  \label{tab:gcp_queue_latency}
  %\end{table}
\end{subfigure}
  
  %\vspace{-1em}
  \caption{Function invocations with serverless queues.}
  %\vspace{-1em}
\end{figure}

\noindent\textbf{Latency.}
We measure the end-to-end latency by triggering an empty
a function that returns a dummy result to the user with a TCP connection.
We consider the best-case scenario of warm invocations with a cached connection to the same client.
The median round-trip latency to the client was 864 \si\microsecond.
In addition to queues, we measure direct function invocations to estimate
the potential of user-side request batching without cloud proxies,
and present AWS and GCP results in Tables~\ref{tab:queue_latency} and ~\ref{tab:gcp_queue_latency}.
Surprisingly, the FIFO queue achieves the lowest latency and is faster than a direct Lambda invocation.
Thus, offloading requests using SQS-based invocation comes with approximately 20ms of overhead.
However, the ordered PubSub subscription is slower than the direct cloud function invocation and unordered subscription, adding over 170 ms of overhead.

\noindent\textbf{Throughput.}
Here, we verify how well queues perform with batching and high throughput loads.
The queue triggers a function that establishes a connection to the client,
and the client measures the median throughput across 10 seconds
(Figure~\ref{fig:faaskeeper_performance_read}).
FIFO queues saturate at the level of a hundred requests per second.
Meanwhile, DynamoDB and standard SQS experience huge variances, leading
to message accumulation and bursts of large message batches.
Thus, we cannot expect to achieve higher utilization in \toolname{}
with a state-of-the-art cloud-native queue, even with ideal pipelining and low-latency storage.
However, we can assign one queue per user, which helps to alleviate scalability concerns partially.

\noindent\textbf{Cost.}
SQS messages are billed in 64 kB increments, and 1 million of them costs \$0.5.
DynamoDB write units are billed in 1 kB increments, and 1 million of them costs \$1.25.
Thus, processing requests via SQS is 160x cheaper than with DynamoDB streams.

\begin{summaryBox}{boxOliveGreen}
SQS provides ordering with cost-efficient invocations.
Nevertheless, it could be the bottleneck for individual clients.
\end{summaryBox}

\begin{figure}[t!]
	\centering
  \includegraphics[width=\linewidth]{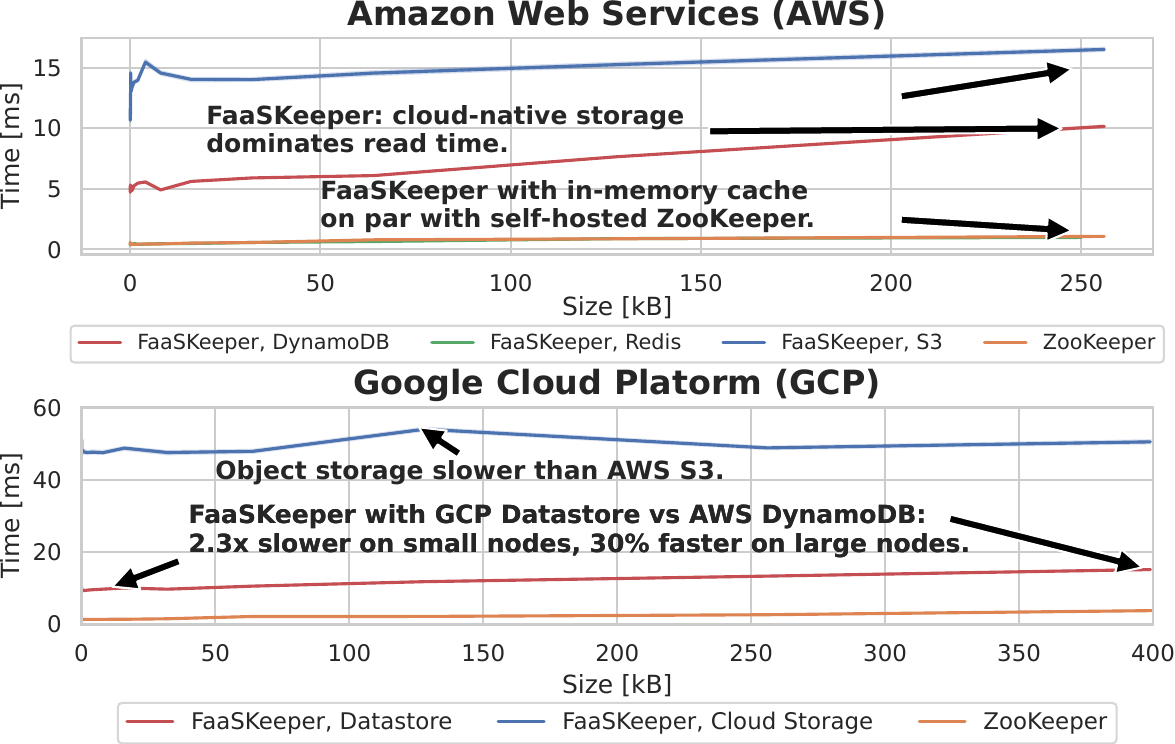}
  \caption{
    Read operations in \toolname{} and ZooKeeper.
  }
  \label{fig:faaskeeper_read}
\end{figure}

\subsection{FaaSKeeper vs ZooKeeper}
\label{sec:evaluation_fk_zk}

We evaluate \toolname{} and compare against ZooKeeper in four domains: read performance, write latency, 
service monitoring, and cost trade-offs.

\subsubsection{Read Operations}
\label{sec:evaluation_reads}
ZooKeeper is designed for efficient read operations,
and our \toolname{} must offer the same.
We evaluate the \code{get\_data} operation that
retrieves a ZooKeeper node, timing the retrieval on the user side.
On AWS, we evaluate S3, DynamoDB and Redis (\code{t3.small} VM) as the user data store.
%and compare \toolname{} against ZooKeeper.
%
On GCP, we evaluate Cloud Storage and Datastore.
We repeat the measurements 100 times for each node size and present results in
Figure~\ref{fig:faaskeeper_read}.
We compare \toolname{} against ZooKeeper, placing the benchmarking client in the same
cloud region zone as one of ZooKeeper's nodes.

Hybrid storage distributes nodes between both storage options (Section~\ref{sec:faaskeeper_cloud_data_storage}),
allowing us to benefit from the low latency of DynamoDB on small nodes while placing large user
data to S3.
This avoids the cost explosion as reading 128 kB data from DynamoDB is 20x more expensive than S3.
ZooKeeper offers much lower latency as it serves data from memory over
a warm TCP connection: \toolname{} matches its performance with an in-memory store.

Sorting results, watches, and deserialization in the client library
adds between 1.9 and 2.5\% overhead in our Python implementation.

\begin{summaryBox}{boxOliveGreen}
\toolname{} offers fast reads whose performance is bounded by the
latency and throughput of the underlying cloud storage,
with a stable cost proportional to workload.
\end{summaryBox}

%\vspace{-1em}
\subsubsection{Write Operations}
\label{sec:evaluation_writes}
\begin{figure}[t!]
	\centering
  \includegraphics[width=\linewidth]{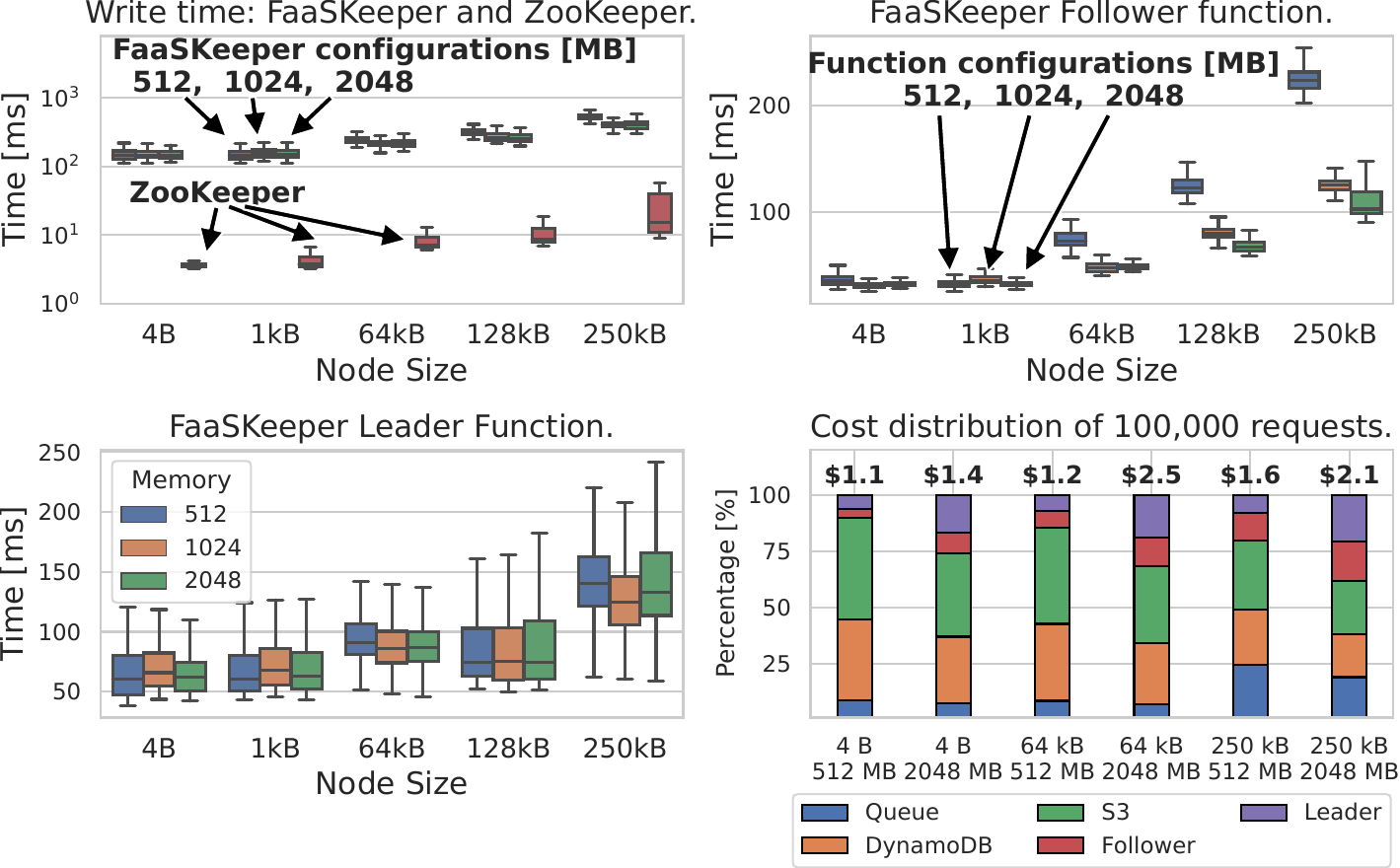}
  \caption{Write operations in \toolname{} and ZooKeeper.
  }
  \label{fig:faaskeeper_write}
\end{figure}

We evaluate the performance and cost of writing in \toolname{} and compare our framework against ZooKeeper.
We measure \code{set\_data} operation that replaces node contents with base64-encoded data of different sizes (Figure~\ref{fig:faaskeeper_write}), pushing to the size limit of 250 kB.
First, we notice that ZooKeeper achieves lower write latency due to the direct connection with a client
and operating on a local state in memory.
The latency in \toolname{} is bounded by the functions and the overheads of queue-based invocations.
Then, we study the execution times of follower and leader functions.
The leader function contributes more to the total write latency, especially on small input sizes, and exhibits a strong variance.
Finally, we look at the writing cost and find that storage operations are responsible for 40-80\% of it, with functions contribution noticeably lower - even though the CPU time of a serverless function is 8x more expensive than in a VM.
Both functions use no more than 100 MB of memory but require larger allocations to increase I/O performance~\cite{copik2020sebs}, leading to increased cost and resource underutilization.

\begin{figure}[t!]
	\centering
   \includegraphics[width=\linewidth]{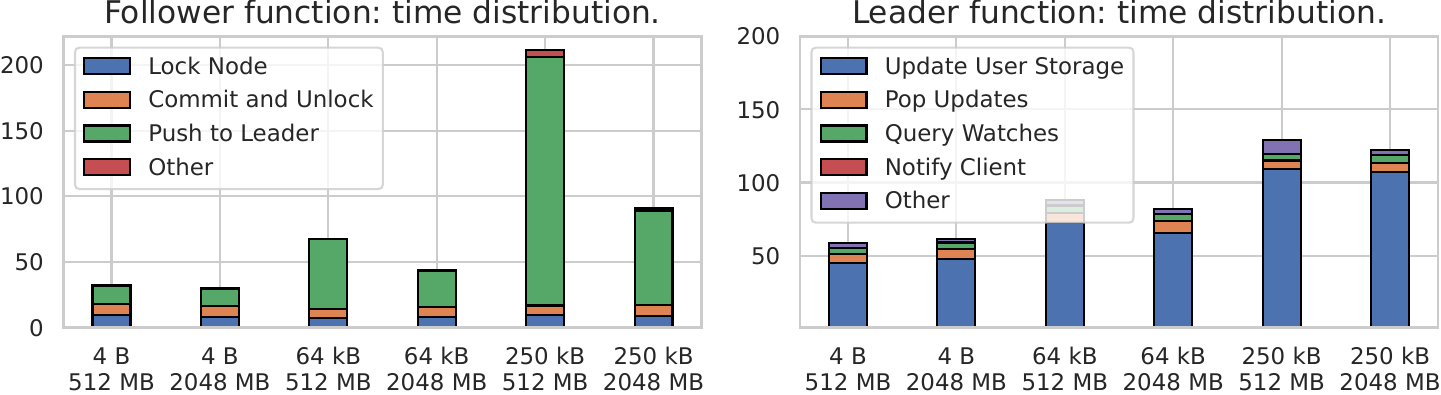}
  \vspace{-1em}
  \caption{Time distribution of \toolname{} functions.}
  \label{fig:faaskeeper_write_decomposition}
\end{figure}

\noindent\textbf{Overhead}
To locate the bottleneck of writing in \toolname{}, we inspect where functions
spend time.
Figure~\ref{fig:faaskeeper_write_decomposition} shows
the impact of synchronization operations is limited, and the runtime of
leader and follower functions are dominated by moving data to queues and storage.
This impacts both the latency and cost, as there is no \emph{yield} operation in serverless - functions waiting on I/O and external services keep consuming resources and accruing costs.
%
%Furthermore, to understand the sources of performance variability, we look at the tail latency of the most important components (Table~\ref{tab:writer_variability}).

\begin{table}[t]
    \centering
    \begin{adjustbox}{max width=\linewidth}
    %\footnotesize
    \small
    \begin{tabular}{lc|r|rrrrr}
      %\toprule
      &\textbf{Follower} & \textbf{Size} & \textbf{Min} & \textbf{p50} & \textbf{p90} & \textbf{p95} & \textbf{p99}\\
      \midrule
\multirow{8}{*}{\rotatebox[origin=c]{90}{\textbf{Follower}}} & \multirow{2}{*}{Total}  &  4B  & 27.29  & 31.81  & 38.55  & 41.88  & 58.78 \\
  &&  250 kB & 30.24  & 102.53  & 142.35  & 163.15  & 183.49 \\
      %\midrule
      %\cline{2-8}
&\multirow{2}{*}{Lock}  &4B  & 7.38  & 8.02  & 9.47  & 12.69  & 26.8 \\
&& 250 kB   & 6.77  & 8.36  & 15.38  & 17.79  & 28.48  \\
      %\cline{2-8}
      %\midrule
&\multirow{2}{*}{Push}  & 4B & 9.65  & 13.35  & 15.55  & 17.28  & 38.15  \\
&& \cellcolor{blue!25}250 kB   & \cellcolor{blue!25}62.73  & \cellcolor{blue!25}72.18  & \cellcolor{blue!25}96.82  & \cellcolor{blue!25}118.62  & \cellcolor{blue!25}148.61 \\
      %\cline{2-8}
      %\midrule
&\multirow{2}{*}{Commit} & 4B  & 7.31  & 7.93  & 9.41  & 11.91  & 26.83  \\
&& 250 kB   & 6.61  & 8.59  & 14.31  & 18.81  & 32.83  \\
      \midrule
\multirow{8}{*}{\rotatebox[origin=c]{90}{\textbf{Leader}}} & \multirow{2}{*}{Total}  & 4B  & 42.02  & 62.16  & 92.01  & 103.65  & 138.28 \\
& &250 kB  & 58.94  & 132.62  & 213.5  & 294.01  & 465.47\\
      %\cline{2-8}
& \multirow{2}{*}{Get Node}  & 4  & 4.67  & 5.09  & 5.68  & 6.92  & 11.83\\
& & 250 kB  & 4.58  & 4.97  & 7.31  & 11.13  & 19.83\\
      %\cline{2-8}
& \multirow{2}{*}{Update Node}& \cellcolor{blue!25}4  & \cellcolor{blue!25}24.4  & \cellcolor{blue!25}42.73  & \cellcolor{blue!25}70.7  & \cellcolor{blue!25}84.94  & \cellcolor{blue!25}118.13\\
& & \cellcolor{blue!25}250 kB   & \cellcolor{blue!25}32.51  & \cellcolor{blue!25}102.07  & \cellcolor{blue!25}183.17  & \cellcolor{blue!25}265.42  & \cellcolor{blue!25}432.92\\
      %\cline{2-8}
& \multirow{2}{*}{Watch Query} & 4  & 3.88  & 4.48  & 5.45  & 7.0  & 28.64\\
& & 250 kB   & 4.68  & 5.13  & 6.76  & 7.59  & 18.38\\
      %\bottomrule
    \end{tabular}
    \end{adjustbox}
  \caption{Variability of functions performance, 2048 MB.}
    %\vspace{-0.5em}
    \label{tab:writer_variability}

\end{table}

\noindent\textbf{Variability}
To understand the sources of performance variability observed in Figure~\ref{fig:faaskeeper_write}, we examine tail latencies of the important operations (Table~\ref{tab:writer_variability}).
We observe significant performance degradation at the tail percentiles when pushing to queue in \emph{follower} and updating S3 nodes in \emph{leader}.
This result aligns with the previous subsection: distributed applications in serverless are particularly affected by inefficient queues and remote storage. 

\begin{figure}[t!]
	\centering
  \includegraphics[width=\linewidth]{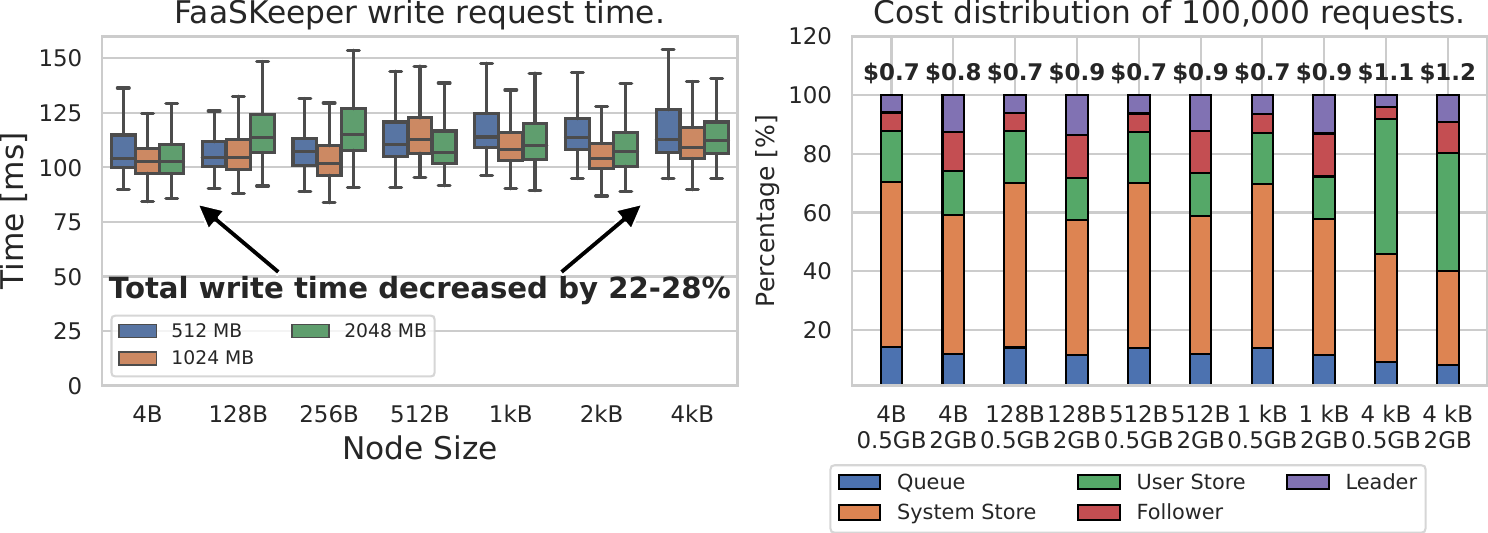}
  \vspace{-1.5em}
  \caption{\toolname{} writes with hybrid storage.
  }
  \label{fig:hybrid_storage}
\end{figure}

\noindent\textbf{Hybrid Storage}
We evaluate the impact of hybrid storage on the node size range typical for ZooKeeper applications (Figure~\ref{fig:hybrid_storage}).
By replacing S3 with DynamoDB for user storage, we improve not only the cost and performance of reading,
but also decrease the write time while keeping costs for infrequent large nodes under control.

\begin{figure}[t!]
	\centering
   \includegraphics[width=\linewidth]{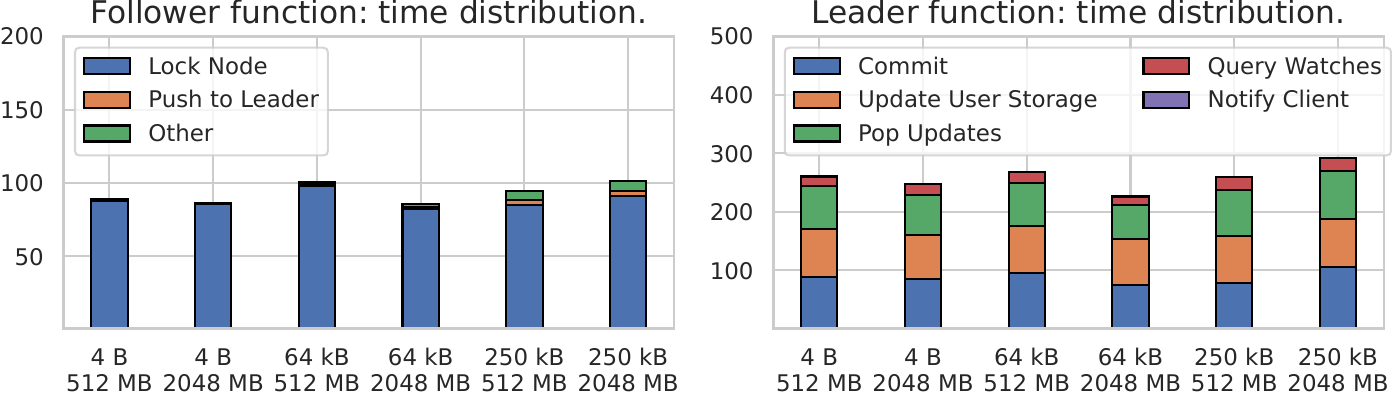}
  \vspace{-1.5em}
  \caption{
    \toolname{} writes on Google Cloud.
  }
  \label{fig:gcp_write}
\end{figure}

% mention vcpu
\noindent\textbf{Google Cloud}
Finally, we evaluate the write performance on Google Cloud (Figure~\ref{fig:gcp_write}).
Compared to AWS, \toolname{} achieves worse performance due to significantly more expensive
synchronization with transactions on key-value storage.
However, the hybrid storage optimization does not apply here since the cost of reading from the NoSQL storage 
is larger than from object storage.

\noindent\textbf{Resource Configuration}
Finally, we explore new configuration options available in serverless.
In Google Cloud, we test the ability to change CPU allocation independently from the memory allocation.
When comparing functions with 512MB memory and 0.33 or 1 virtual CPU, 
we notice performance change of 2-10\%, often favoring the smaller allocation.
However, a smaller CPU allocation translates to a 54-62\% cost decrease.
Applications like FaaSKeeper are I/O-bound and benefit from flexible allocation of CPU resources.
Then, we compare the x86 and ARM instances of AWS Lambda.
There, ARM functions perform better on follower functions but experience significant slowdowns of up to 94\% on the leader function.
Depending on the configuration, ARM functions can decrease costs of follower functions by up to 32\%.

\begin{summaryBox}{boxOliveGreen}
Write operations are limited by data transmission to
queues and object storage, motivating
the need for more efficient queues.
\end{summaryBox}

\subsubsection{Service Monitoring}
\label{sec:evaluation_heartbeat}
\begin{figure}[t!]
	\centering
  \includegraphics[width=\linewidth]{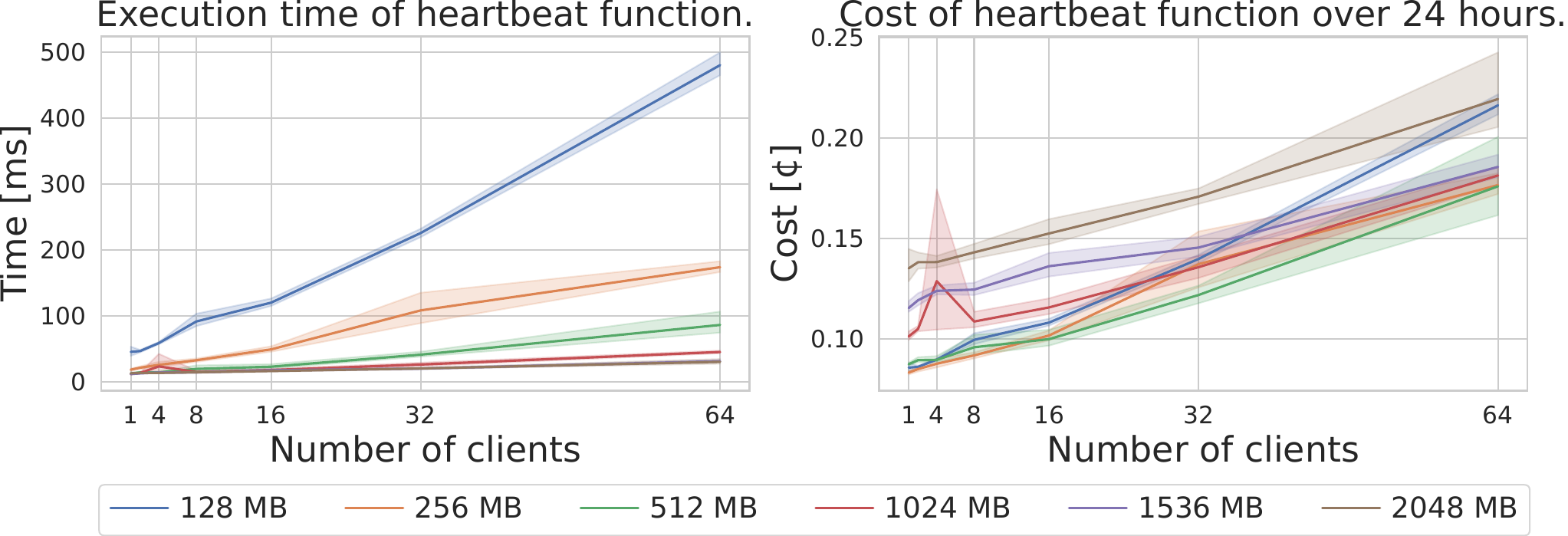}
  \caption{Heartbeat function performance and cost.
  } 
  \label{fig:faaskeeper_heartbeat}
\end{figure}

We estimate the time and resources needed by FaaSKeeper to periodically
launch the \code{heartbeat} function and verify status of clients owning ephemeral nodes.
We present results averaged from 100 invocations in Figure~\ref{fig:faaskeeper_heartbeat}.
Execution time decreases with the allocation, corresponding
with previous findings on serverless I/O~\cite{copik2020sebs,10.5555/3277355.3277369}.

We estimate the cost of monitoring over the entire day,
with the highest available frequency on AWS Lambda of an execution every minute.
The cost of the function is defined by the computation time and the cost of
scanning a DynamoDB table storing the list of users.
With the function taking less than 100ms for most configurations,
the overall allocation time over 24 hours is less than 0.2\% of the entire day.
Thus, even for more frequent invocations and more clients, we offer status
monitoring for a fraction of VM price.

\begin{summaryBox}{boxOliveGreen}
The serverless heartbeat function replaces a persistent VM allocation and achieves the
goal of client monitoring while reducing the resource allocation time by a huge margin.
\end{summaryBox}

%\vspace{-1em}

\begin{figure}[t]
	\centering
  \begin{subfigure}{\linewidth}
    \includegraphics[width=\textwidth]{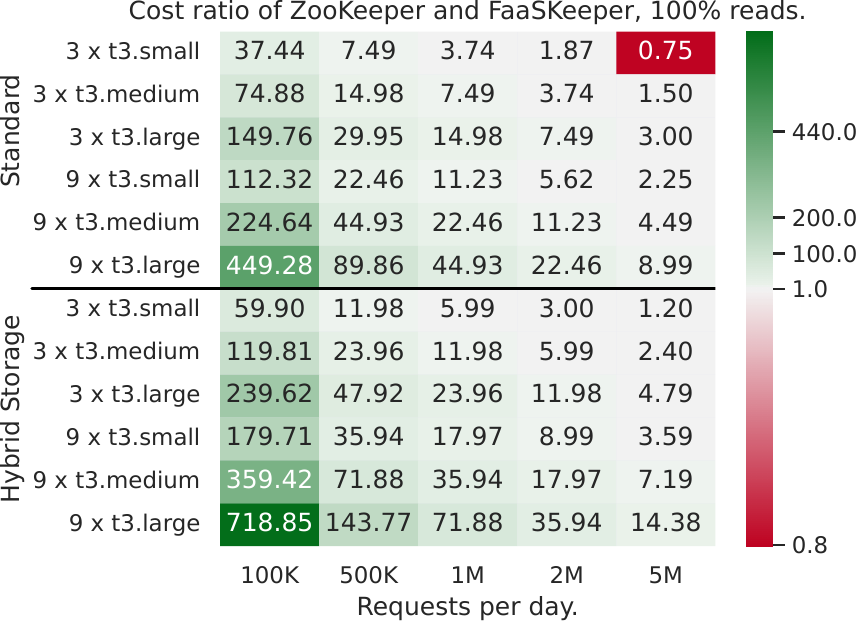}
  \end{subfigure}
  \hfill
  \begin{subfigure}{\linewidth}
    \includegraphics[width=\textwidth]{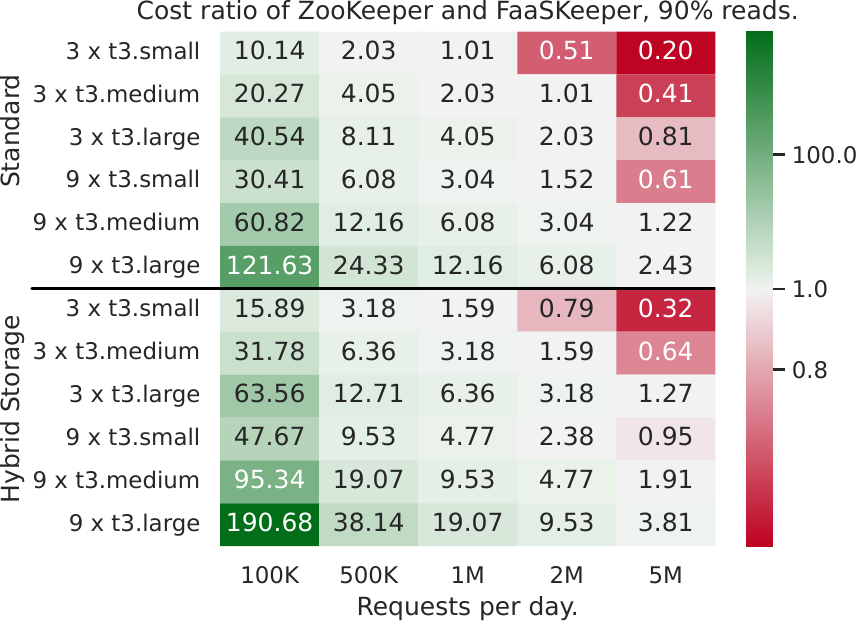}
  \end{subfigure}
  \hfill
  \begin{subfigure}{\linewidth}
    \includegraphics[width=\textwidth]{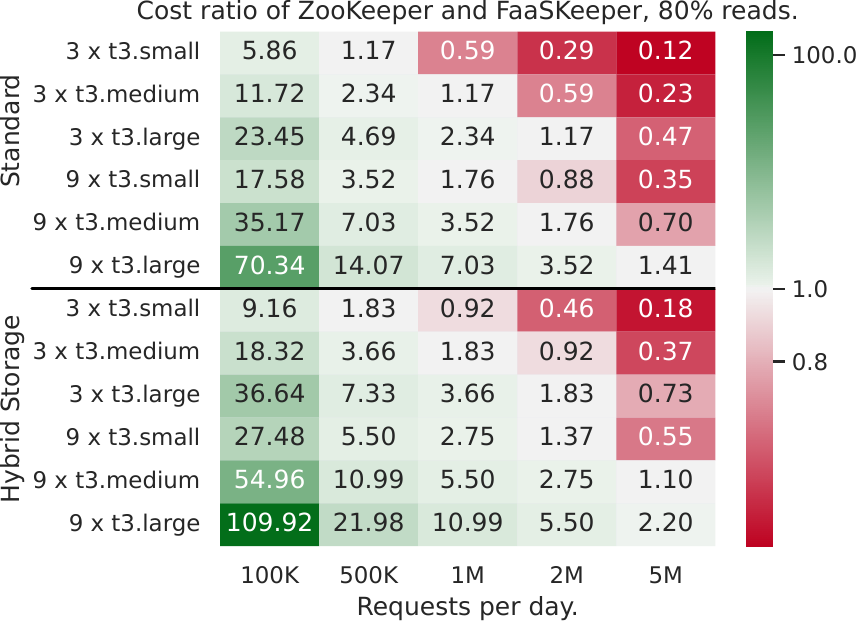}
  \end{subfigure}
  \vspace{-1em}
  \caption{Cost ratio of ZooKeeper and \toolname{}, running a workload mix of 1 kB reads and writes with \texttt{set\_data}.}
  \label{fig:faaskeeper_cost}
\end{figure}

\begin{table}\centering
	\begin{adjustbox}{max width=\linewidth}
  \footnotesize
  \begin{tabular}{llc}
    %\toprule
    Parameter & Description & Value\\
		\midrule
    $W_{S3}(s)$   & Writing data to S3 & $5\cdot10^{-6}$\\
    $R_{S3}(s)$   & Reading data from S3 & $4\cdot10^{-7}$\\
    $W_{DD}(s)$   & Writing data to DynamoDB & $ s\cdot1.25\cdot10^{-6}$\\
    $R_{DD}(s)$   & Reading data from DynamoDB & $ \left \lceil{\frac{s}{4}}\right \rceil \cdot0.25*10^{-6}$\\
    $Q(s)$        & Push to queue& $\left \lceil{\frac{s}{64}}\right \rceil \cdot 0.5\cdot10^{-6}$\\
    $F_{W/D}(s)$  & Execution of follower and leader function. & -\\
	\end{tabular}
\end{adjustbox}
%\vspace{-1em}
\caption{\textbf{Parameters of \toolname{} cost model.}}
\label{tab:faaskeeper_properties_model}
\end{table}
%\vspace{-2em}

\subsubsection{Cost}
\label{sec:faaskeeper_properties}

The most important evaluation compares the price of running an elastic \toolname{} instance to Zookeeper
with standard and hybrid storage on AWS, with x86 functions.
We consider a scenario of 512 MB, with reads and writes to one node of 1 kB, and the optimistic case that we experience no failures and, therefore, no retries.

\noindent\textbf{\toolname{}}
We focus on read and write operations of $s$ kilobytes,
as the daily monitoring costs are low.
Watch and heartbeat functions add charges only when notifications and ephemeral nodes
are used.
We model the cost of modifying node data (\texttt{set\_data} in ZooKeeper), and summarize model parameters in Table~\ref{tab:faaskeeper_properties_model}.

\noindent\emph{Reading.}
The cost of operation is limited to storage access.
\begin{equation*}
  \textproc{Cost}_{R} = R_{S3}(s)
\end{equation*}
A workload of 100,000 read operations costs \$0.04.

\noindent\emph{Writing.}
The cost of writing is separated into computing and storing data:
two queue operations, function executions,
synchronization in the \code{follower} and \code{leader},
and writing data to the user store.
\begin{equation*}
  \textproc{Cost}_{W} = 2\cdot Q(s) + 3 \cdot W_{DD}(1) + R_{DD}(1) +  W_{S3}(s) + F_{W} + F_{D}
\end{equation*}
A workload of 100,000 write operations costs \$1.12.
With hybrid storage, the cost of user storage write $W_{S3}(s)$ becomes $W_{DD}(s)$
There, a workload of 100,000 write operations costs \$0.72.

\noindent\emph{Storage.}
The databases and queues do not generate any inactivity charges except for retaining data. % in the cloud.
Storing user data in S3 with \toolname{} is 3.47x cheaper than storing the same data in the block
storage \code{gp3} attached to the EC2 virtual machines hosting ZooKeeper.
The hybrid storage incurs higher costs, as retaining data in DynamoDB is 3.125x more expensive than block storage.
However, the size of ZooKeeper data is not high as nodes are usually small, and data access costs dominate the long-term storage.

\noindent\textbf{ZooKeeper}
The cost is constant and includes the cost of a persistent allocation of virtual
machines.
The smallest number of virtual machines is three.
However, a single machine with an attached EBS block storage has an annual durability of 99.9\%. 
To match the annual durability of S3 used as the user store in \toolname{} (11 9's), the ZooKeeper ensemble requires nine machines.
Depending on the VM selection, the daily cost changes from \$0.5 on \code{t3.small},
through \$1 on the \code{t3.medium} used for our experiments, up to \$2 on \code{t3.large}.
Additionally, the machines must be provisioned with block storage to store OS, ZooKeeper, and user data.
20GB of storage adds a monthly cost of between \$4.8 (3 VMs) and \$14.4 (9 VMs).

\noindent\textbf{Comparison}
We compare ZooKeeper's cost against \toolname{} with different read--to--write scenarios,
using 1kB writes and functions configured with 512 MB of memory,
and present results in Figure~\ref{fig:faaskeeper_cost}.
In high--read--to--write scenarios for which ZooKeeper has been designed, \toolname{} can process between 1 and $3.75$ million requests daily before the costs equal the smallest possible ZooKeeper deployment.
With hybrid storage, this number grows to $5.99$ million daily read requests.
Since many user nodes do not contain large amounts of data,
\toolname{} can handle the daily traffic
of hundreds of thousands of requests while providing lower costs than ZooKeeper.
Contrary to the standard ZooKeeper instance, the serverless design allows us to
limit expensive computing time to processing writes only.
Furthermore, we can \emph{shut down} the processing components while not losing any data:
the heartbeat function is suspended after the deregistration of the last client,
and the only charges come from the durable storage of the system and user data.
%

%-------------------------------------------------------------------------------

\section{Building Serverless Services}

\label{sec:discussion}

In the following section, we address the primary challenges encountered during the development of \toolname{} and highlight the current limitations of serverless technology. 
We compile a set of requirements cloud providers could easily support and pave the way for future improvements. 
They would make complex serverless systems more efficient and performant, simplifying their implementation and increasing adoption.
In particular, they would allow \toolname{} to match ZooKeeper's performance when using off-the-shelf cloud services.
We finally discuss how well these requirements are supported in research and emerging cloud architectures.

\subsection{Areas of improvement}
Using the lessons learned while creating \toolname{}, we propose a list of requirements for serverless environments that would allow complex services to flourish.
However, the rationale behind these requirements is not limited to our use case, and will improve
other applications, such as microservices~\cite{10.1145/3476886.3477510}
and serverless ML~\cite{10.1145/3357223.3362711,10.1145/3448016.3459240}.

\noindent \textbf{Requirement \#1: Fast invocations.}
Invocation overheads dominate the execution time of short-running functions~\cite{copik2020sebs}
and prohibit FaaS processing with performance comparable to non-serverless applications that can use
a direct RPC call over a TCP connection.
ZooKeeper often requires multiple round trips to finish an operation, and when each one takes milliseconds rather than microseconds, the overheads quickly accrue, as seen in Figure~\ref{fig:faaskeeper_write}.

\noindent \textbf{Requirement \#2: Exception handling.}
The user cannot control asynchronous function invocations %requiring notifications of completed operations 
(Section~\ref{sec:faaskeeper_design_distributor}).
We envision this should be solved via user-defined \emph{exception handlers},
allowing for easier and more efficient error handling.

\noindent \textbf{Requirement \#3: Synchronization primitives.}
To efficiently implement distributed applications, serverless needs synchronization, such as
locks and atomics (Section~\ref{sec:faaskeeper_serverless_components}).
In practice, sub-millisecond latency is needed, like the one offered by in-memory storage.

\noindent \textbf{Requirement \#4: FIFO Queues.}
Serverless functions require queues to support the ordering and
reliability of invocations (Section~\ref{sec:faaskeeper_design_distributor}).
However, queues that use discrete batches prevent efficient stream processing
with serverless functions.
Instead, functions should continuously poll for new items in the queue
to keep the pipeline saturated.
Furthermore, they can be significantly slower than regular invocations (Section~\ref{sec:evaluation_queues}).

\noindent \textbf{Requirement \#5: Statefulness.}
While stateless functions are sufficient for many use cases, \emph{stateful} functions are necessary
to efficiently process requests that depend on each other (Section~\ref{sec:faaskeeper_design_distributor}).
FaaS should support a reliable and low-latency function state.

\noindent \textbf{Requirement \#6: Partial updates.}
To increase the efficiency of write operations, object storage could support partial updates
where data is written at a user-defined offset to the specified object, avoiding the need for the read-update-write process (Section~\ref{sec:faaskeeper_design_distributor}).

\noindent \textbf{Requirement \#7: Outbound channels.} While the trigger system provides \emph{inbound}
communication, functions lack an ordered, push-based, and fast \emph{outbound}
communication channel with acknowledgment of delivery.
Cloud queues are an order of magnitude slower than a TCP connection and do not validate that the recipient read the message.
Such a channel would significantly simplify the design of serverless services such as \toolname{} (Section~\ref{sec:faaskeeper_design_distributor}).

\noindent \textbf{Requirement \#8: Fast serverless storage.}
ZooKeeper is a data-intensive system focused on fast read operations.
In \toolname{}, in-memory storage could deliver competitive performance, but it is not available
as serverless and cloud-native service (Section~\ref{sec:evaluation_reads}).

\noindent \textbf{Requirement \#9: Decoupling I/O and compute.}
\toolname{} functions spend most of the time waiting on requests
to cloud services -- increasing CPU allocation alone has little effect on performance (Section~\ref{sec:evaluation_writes}).
Furthermore, queue batching (R4) prevents effective handling of many requests within a single function.
Instead, functions should be swapped out during idle periods to free up resources.
While this is a fundamental change, improved I/O management would decrease user costs
and allow cloud providers to increase utilization with a larger degree
of oversubscription.

\subsection{Discussion}

\textbf{Can serverless systems support our requirements?}
We specify \reqCount{} requirements to define features missing in cloud-native FaaS systems that are necessary for distributed, stateful, and scalable applications.
%
% latency - Berkeley 2019, Lopez 2021
% fixme - cite stateful and storage
The requirements align with the major serverless challenges~\cite{DBLP:journals/corr/abs-1902-03383,lopez2021serverless}
and are supported in research FaaS platforms.
Emerging systems provide microsecond-scale invocation latency~\cite{nightcore,copik2021rfaas}
and I/O separation from functions~\cite{10.1145/3620678.3624648}.
New storage systems satisfy the latency, consistency, and flexibility requirements
of functions~\cite{216007,10.5555/3291168.3291200,10.1145/3357223.3362723,10.14778/3407790.3407836},
including serverless in-memory caches~\cite{246184,10.14778/3587136.3587139}.
Furthermore, stateful serverless is becoming the new norm in clouds~\cite{10.1145/3361525.3361535,10.1145/3477132.3483541,254432,praas,258880}.
Finally, we note that research systems can support many of our requirements
already: Cloudburst (R1, R5)~\cite{10.14778/3407790.3407836},
PraaS (R1, R5, R7, R9)~\cite{praas}, Boki (R3-R5)~\cite{10.1145/3477132.3483541}.

%primitives - custom lithops java

\textbf{What are the design trade-offs of \toolname{}?}
\toolname{} achieves elastic scaling and a serverless price model by accepting the increased latency of FaaS
systems.
However, performance overheads are isolated to specific services
and their impact will decrease with the adoption
of more efficient serverless platforms.
\toolname{} can match ZooKeeper's read performance by incorporating an
in-memory database, but these are unavailable as a cloud-native serverless service and require third-party solutions~\cite{246184,upstash}.
The increased processing time of write requests is caused primarily
by performance variations of cloud queues and object storage.

%-------------------------------------------------------------------------------

%-------------------------------------------------------------------------------
\section{Related Work}

\emph{Serverless for Storage}
Wang et al.~\cite{246184} use functions for elastic in-memory cache.
Boki provides stateful serverless on shared logs~\cite{10.1145/3477132.3483541}.
DynamoDB is used in transactional workflows with locks in Beldi~\cite{258880}
and in a fault-tolerance shim AFT~\cite{10.1145/3342195.3387535}.
In contrast, \toolname{} is designed as a service and not a backend for serverless functions.
We offer coordination for general-purpose applications while optimizing resource allocation.

\textlambda{}FS implements a serverless metadata layer of a distributed file system~\cite{10.1145/3623278.3624765}.
Similarly to \toolname{}, functions operate on top of strongly consistent data storage.
However, the implementation of read operations is different in both systems.
In \textlambda{}FS, metadata reading is handled by functions that use caching to avoid reaching to the data store.
In \toolname{}, functions are removed entirely from the reading path.
Finally, \textlambda{}FS also identified some of the requirements for efficient serverless applications (Section~\ref{sec:discussion}),
such as fast invocations that bypass the slow interface of HTTP requests.

\noindent\emph{Elastic Storage}
Cloud-native storage is known for elastic implementations
that scales with changes in workload~\cite{10.1145/2063576.2063973}.
Examples of reconfiguration controllers include
a reactive model using CPU utilization
as the primary metric for scaling of
in the Hadoop Distributed File System\cite{10.1145/1809049.1809051},
a feedforward and feedback controller for key-value storage to resize the service and 
minimize SLO violations~\cite{10.1145/2494621.2494630},
a workload-aware heterogenous reconfiguration engine for HBase~\cite{10.1145/2465351.2465370},
a workload predictor with cost-aware reconfiguration~\cite{234950},
latency monitoring and forecasting in database-agnostic replication techniques~\cite{183067}.
PolarDB is an example of a disaggregated database that offers a serverless
billing model~\cite{10.1145/3448016.3457560}.
However, ZooKeeper requires autoscaling procedures that integrate the state ordering guarantees.
\toolname{} achieves that by using the auto-provisioning of serverless functions and databases.

\noindent\emph{ZooKeeper}
Other works explored different approaches to ZooKeeper's performance.
Stewart et al.~\cite{180172} replicated data on multiple servers to provide predictable
access latencies.
Distler et al.~\cite{10.1145/2741948.2741954} introduced ZooKeeper extensions to optimize
coordination patterns by performing additional work on the server.
Shen et al.~\cite{10.1145/2987550.2987561} proposed live migration for geographical reconfiguration.
The performance of ZooKeeper has been improved with
hardware implementations, using FPGAs~\cite{194954} and offloading to network adapters~\cite{stalder2020zoo}.
with PsPIN~\cite{di2020pspin}.

Other systems provide similar semantics guarantees and semantics as ZooKeeper.
Shi et al.~\cite{6968766} presented Giraffe, a coordination service providing higher write performance
and improved availability over ZooKeeper while keeping the same guarantees.
Halalai et al.~\cite{6983381} presented ZooFence, an automatic service partitioning
built on top of vanilla ZooKeeper instances.
Their solution achieves higher scalability than a standard ZooKeeper instance while
upholding all consistency guarantees.
Schiekofer et al.~\cite{8023134} presented Agora, a ZooKeeper-like system that achieves higher
throughput by dynamically splitting data into independent and parallel partitions.
ZooKeeper data is split into partitions updated independently on each server.
To achieve the usual consistency guarantees and prevent clients from observing an incorrect sequence
of updates, read requests for a given partition are stalled when the client has seen newer data
on another partition.
Clients can use \emph{fast} reads with weaker consistency that return the stale partition
state and ignore potential inter-partitional dependencies.
\toolname{} implements the parallelization across serverless workers without the need for data decomposition.
Furthermore, our design allows for more flexible resource allocation thanks to data and compute
disaggregation.

%-------------------------------------------------------------------------------

%-------------------------------------------------------------------------------
\section{Conclusions}

As the tools and mechanisms of cloud computing adapt to the needs of an
ever-growing FaaS landscape, creating a powerful, fast, and efficient
serverless application is becoming possible.
In this work, we present \toolname{}, a cloud-native and serverless coordination service
offering the same consistency model and interface as Zookeeper.
\toolname{} allows for an elastic deployment that matches system activity,
reducing the cost of some configurations by a factor of up to 719x.
We discuss the lessons learned in creating \toolname{}, and identify \reqCount{} requirements that
clouds should fulfill to ensure functionality and performance.

%-------------------------------------------------------------------------------

\begin{acks}
This project received funding from EuroHPC-JU under grant agreements DEEP-SEA, No 955606 and RED-SEA, No 955776.
We thank Amazon Web Services for supporting this research with credits through the AWS Cloud Credit for Research, 
and Google Cloud Platform through the Google Cloud Research Credits program with the award GCP19980904.
\end{acks}

\appendix

\section{ZooKeeper}
\label{sec:appendix_zookeeper}
Below we summarize the provided consistency requirements~\cite{10.5555/1855840.1855851,10.5555/2904421,zookeeperDocs} briefly,
considering the case of $M$ clients $C_{1}, \dots, C_{M}$ using a ZooKeeper instance
consisting of $N$ servers $S_{1}, \dots, S_{N}$.

%\setlength{\parindent}{0em}
%\setlength{\parskip}{0em}
% Sources:
% ZK docs
% Atomicity : Updates either succeed or fail -- there are no partial results.
\paragraph{\nodeColor{Z1}{black}{ellipse} Atomicity}
Write requests never lead to partial results. They are accepted and persistently committed
by ZooKeeper or they fail.
%
% Sources:
% ZK docs
% Sequential Consistency - Updates from a client will be applied in the order that they were sent.
\paragraph{\nodeColor{Z2}{black}{ellipse} Linearized Writes}
If a client $C_{i}$ sends update request $u$ before request $v$, and both
are accepted,
%resulting in transactions with \emph{zxid} $u'$ and $v'$, respectively,
then it must hold that u "happens before" v, i.e., $u < v$.
The guarantee holds for a single session.
When clients $C_{i}$ and $C_{j}$ send requests $u_{1}, u_{2}, \dots$ and $v_{1}, v_{2}, \dots$,
respectively,
%the ordering between any $u_{i}'$ and any $v_{j}'$ is not defined.
the ordering between any $u_{i}$ and $v_{j}$ is not defined.
%If any client $C_{i}$ sends update request $u$ before another request $v$, and both requests
%are successfully applied,
%%resulting in transactions with \emph{zxid} $u'$ and $v'$, respectively,
%it must hold that u "happens before" v, i.e., $u < v$.
%%
%The guarantee holds within a single session.
%%
%When clients $C_{i}$ and $C_{j}$ send update requests $u_{1}, u_{2}, \dots$ and $v_{1}, v_{2}, \dots$,
%respectively,
%%the ordering between any $u_{i}'$ and any $v_{j}'$ is not defined.
%the ordering between any $u_{i}$ and any $v_{j}$ is not defined.

% Sources:
% ZK docs
% Single System Image - A client will see the same view of the service regardless of the server that it connects to.
% Reliability - Once an update has been applied, it will persist from that time forward until a client overwrites the update.
% Timeliness - The clients view of the system is guaranteed to be up-to-date within a certain time bound.

\paragraph{\nodeColor{Z3}{black}{ellipse} Single and Reliable System Image}

The order of successful updates is visible as identical to every client:
for any updates $u$ and $v$,
if a client $C$
connected to a server $S$
observes that $u < v$,
it must hold that $u < v$ for any client $C'$ connected to any server $S'$.
Furthermore if a client $C$ observes node $Z$ with version $V$, it cannot later
see the node $Z$ with version $V'$ such that $V' < V$, even if session mechanism
switched servers due to failure or network outage.
Each view of the system will become up-to-date after bounded time,
or a disconnection notification will be delivered (\emph{timeliness}).
Accepted updates are never rolled back.
%The order of successful updates is visible as identical from all servers: 
%for any two updates $u$ and $v$,
%if a client $C$
%connected to a server $S$
%observes that $u < v$,
%it must hold that $u < v$ for any client $C'$ connected to any server $S'$.
%%
%Furthermore if a client $C$ observes node $Z$ with version $V$, it cannot later
%see the node $Z$ with version $V'$ such that $V' < V$, even if session mechanism
%switched servers due to failure or network outage.
%%
%Each view of the system will become up-to-date after bounded time,
%or a disconnection notification will be delivered (\emph{timeliness}).
%%
%Accepted updates are never rolled back.

% Sources:
% ZK docs
% With regard to watches, ZooKeeper maintains these guarantees:
% Watches are ordered with respect to other events, other watches, and asynchronous replies. The ZooKeeper client libraries ensures that everything is dispatched in order.
% A client will see a watch event for a znode it is watching before seeing the new data that corresponds to that znode.
% The order of watch events from ZooKeeper corresponds to the order of the updates as seen by the ZooKeeper service.
\paragraph{\nodeColor{Z4}{black}{ellipse} Ordered Notifications}

Watch notifications are delivered in the order of updates that triggered them.
Their ordering with respect to other notifications and writes must be preserved.
If an update $u$ triggers a watch notification for a client $C$, the client must observe
the notification before seeing any data touched by transaction $v$ such that $u < v$.
In particular, if a client $C$ has a watch registered on any node $Z$ with version $V$,
it will receive watch notification before seeing any data associated with node $Z$ with version
$V'$ such that $V < V'$.
The property outlined above is global, i.e., it affects all changes preceded by the notification,
not only changes related to watches registered by the client.

\section{FaaSKeeper Consistency Model}
\label{sec:appendix_faaskeeper}

\paragraph{\nodeColor{Z1}{black}{ellipse} Atomicity}
The updates in the system storage are performed in a single operation on the key-value
storage that is guaranteed to be atomic.
The operation results are propagated to the leader queue before the commit.
The queue triggers the leader function and retries it upon failure, guaranteeing the eventual propagation of changes to all data replicas.
Since the leader function verifies node status before propagating changes 
(\numdingDark{1} in Alg.~\ref{alg:distributor_function}), incorrect operations do not affect the system.

\paragraph{\nodeColor{Z2}{black}{ellipse} Linearized Writes}
Updates are processed in a FIFO order by the \emph{follower} function.
The queue guarantees that only a single follower instance can be active at a time,
and the function is not allowed to reorder any two requests unless they come from a different session.
Therefore, any two update requests $u$, $v$ in the same session cannot be assigned a timestamp
value such that $u \ge v$.
The single \emph{leader} instance guarantees that clients reading from user data
never observe $v$ before $u$.
Different sessions can use different queues and see their respective requests be reordered, which
conforms to ZooKeeper's undefined ordering of requests between clients. 

\paragraph{\nodeColor{Z3}{black}{ellipse} Single and Reliable System Image}
Nodes are stored in a cloud storage with automatic replication
and a strongly consistent read must always return the newest data.
Thus, if a client $C$ observes updates $u, v$ such that $u < v$, all other
clients must read either the same or newer data.
Furthermore, strongly consistent reads prevent clients from observing an order of updates
$V, V', V$.
%Since all nodes are stored in a cloud storage with automatic replication,
%a strongly consistent read following a write must return the newest data.
%%
%Thus, if any client $C$ observes updates $u, v$ such that $u < v$, all other clients must
%read either the same or newer data.
%%
%Furthermore, strongly consistent reads prevent clients from observing an order of updates
%$V, V', V$ for the same node.

\paragraph{\nodeColor{Z4}{black}{ellipse} Ordered Notifications}
\toolname{} guarantees that transactions with timestamp $v$ are not visible before receiving all
notifications corresponding to updates $v'$ such that $v' < v$.
When a read returns a node with timestamp $v$, it is first compared with the $MRD$ value of the current session.
If $v < MRD$, then by the transitive property of the total order, any pending watch notifications
must be newer than $v$, and data is safe to read.
Otherwise, there are two possible situations:
(a) a watch notification relevant to the client was active but not yet delivered
(\numdingDark{4} in Alg.~\ref{alg:distributor_function})
before storing $v$ (\numdingDark{2} there),
and (b) no relevant watch notifications are being processed.

In the former case, if a transaction $v'$ triggers watch $w$,
it is added to the $epoch$ counter before committing $v$,
Thus, for each transaction following $v'$, watch $w$ must be included in the $epoch$ unless the
notification is delivered to each client (\numdingDark{6}).
This prevents the client from seeing transaction $v$ unless watch $w$ is notified.
In the latter case, the client library releases the data immediately because watch $w$ is not
present in the $epoch$.

\bibliographystyle{ACM-Reference-Format}
\bibliography{aws,zookeeper,serverless,cloud}

%%%%%%%%%%%%%%%%%%%%%%%%%%%%%%%%%%%%%%%%%%%%%%%%%%%%%%%%%%%%%%%%%%%%%%%%%%%%%%%%
\end{document}